\documentclass[acmsmall]{acmart}
\usepackage{algorithmic}
\usepackage{tabularx}
\usepackage{graphicx}
\usepackage{tcolorbox}
\usepackage{hyperref}
\usepackage{textcomp}
\usepackage{array}
\usepackage{bmpsize}
\usepackage{lipsum}
\usepackage{booktabs}
\usepackage{colortbl}  % For coloring rows and columns
\usepackage{xcolor}    % For alternating row colors
\usepackage{multirow}  % For multirow cells

\AtBeginDocument{%
  }

\setcopyright{acmlicensed}
\copyrightyear{2025}
\acmYear{2025}
\acmDOI{XXXXXXX.XXXXXXX}

\begin{document}

%%
%% The "title" command has an optional parameter,
%% allowing the author to define a "short title" to be used in page headers.
\title{A Survey on Vulnerability Prioritization: Taxonomy, Metrics, and Research Challenges}

%%
%% The "author" command and its associated commands are used to define
%% the authors and their affiliations.
%% Of note is the shared affiliation of the first two authors, and the
%% "authornote" and "authornotemark" commands
%% used to denote shared contribution to the research.

\author{Yuning Jiang}
\email{yuning\_j@nus.edu.sg}
\orcid{0000-0003-4791-8452}
\affiliation{%
  \institution{National University of Singapore}
  \country{Singapore}
}
  
\author{Nay Oo}
\email{nay.oo@ncs.com.sg}
\affiliation{%
 \institution{NCS Cyber Special Ops R\&D}
  \country{Singapore}}

  \author{Qiaoran Meng}
\email{qiaoran@nus.edu.sg}
\affiliation{%
  \institution{National University of Singapore}
  \country{Singapore}}

\author{Hoon Wei Lim}
\email{hoonwei.lim@ncs.com.sg}
\affiliation{%
 \institution{NCS Cyber Special Ops R\&D}
  \country{Singapore}}

\author{Biplab Sikdar}
\email{bsikdar@nus.edu.sg}
\affiliation{%
  \institution{National University of Singapore}
  \country{Singapore}}

%%
%% By default, the full list of authors will be used in the page
%% headers. Often, this list is too long, and will overlap
%% other information printed in the page headers. This command allows
%% the author to define a more concise list
%% of authors' names for this purpose.
\renewcommand{\shortauthors}{Jiang et al.}

%%
%% The abstract is a short summary of the work to be presented in the
%% article.
\begin{abstract}

In today's highly interconnected digital landscape, safeguarding complex infrastructures against cyber threats has become increasingly challenging due to the exponential growth in the number and complexity of vulnerabilities. Resource constraints necessitate effective vulnerability prioritization strategies, focusing efforts on the most critical risks. This paper presents a systematic literature review of 82 studies, introducing a novel taxonomy that categorizes metrics into severity, exploitability, contextual factors, predictive indicators, and aggregation methods. Our analysis reveals significant gaps in existing approaches and challenges with multi-domain applicability. By emphasizing the need for dynamic, context-aware metrics and scalable solutions, we provide actionable insights to bridge the gap between research and real-world applications. This work contributes to the field by offering a comprehensive framework for evaluating vulnerability prioritization methodologies and setting a research agenda to advance the state of practice.

\end{abstract}

%%
%% The code below is generated by the tool at http://dl.acm.org/ccs.cfm.
%% Please copy and paste the code instead of the example below.
%%


%%
%% Keywords. The author(s) should pick words that accurately describe
%% the work being presented. Separate the keywords with commas.

%\received{20 February 2007}
%\received[revised]{12 March 2009}
%\received[accepted]{5 June 2009}

%%
%% This command processes the author and affiliation and title
%% information and builds the first part of the formatted document.
\maketitle

\section{Introduction}

In today's interconnected and digitized world, securing complex systems is paramount for protecting sensitive information and maintaining critical operations. As organizations increasingly rely on technology, the number and complexity of vulnerabilities within software and hardware systems have grown exponentially \cite{iannone2022secret}. Effective and timely remediation strategies are essential to mitigate potential cyber threats. However, resource constraints and the overwhelming volume of vulnerabilities make it impractical to address every potential threat comprehensively \cite{le2022survey, mehri2023automated}. This limitation necessitates the prioritization of vulnerabilities based on their risk.

Cybersecurity risk metrics play a crucial role in this prioritization, enabling organizations to focus defensive measures on the most critical vulnerabilities. These metrics aim to assess the technical severity of vulnerabilities and contextual factors that may influence their impact \cite{cheimonidis2024dynamic, hore2023deep}. Despite significant progress in this field, the increasing complexity of modern systems and networks calls for a systematic evaluation of how these metrics address both technical and contextual aspects of risk \cite{zeng2023illation, yan2022cyber}.

In this paper, we aim to understand the evolving landscape of vulnerability prioritization. Our investigation is structured around the following research questions:

\begin{itemize}
\item[\textbf{RQ1}:] \textit{Which metrics are commonly used to prioritize vulnerabilities, and how do they capture the severity and exploitability of vulnerabilities?}

\item[\textbf{RQ2}:] \textit{What are the predominant methodologies used for prioritizing vulnerabilities, and how do they address the challenges of effective vulnerability management?}

\item[\textbf{RQ3}:] \textit{What are the key challenges in current vulnerability prioritization approaches, and what trends and future directions are emerging in the field?}
\end{itemize}

By addressing these research questions, we aim to provide a comprehensive analysis of how vulnerabilities are prioritized in practice and how future research can address existing limitations. \textbf{RQ1} investigates the commonly used metrics and evaluates their effectiveness in capturing both technical severity and contextual exploitability. \textbf{RQ2} examines the methodologies developed to prioritize vulnerabilities, highlighting their strengths, limitations, and practical applications across various environments. \textbf{RQ3} identifies the key challenges faced by current approaches and uncovers emerging trends and innovative techniques that could shape the future of vulnerability prioritization.

Using a systematic literature review, we collected and analyzed 82 studies from reputable databases such as the ACM Digital Library, IEEE Xplore, Scopus, and Google Scholar. From this analysis, we introduce a novel taxonomy of vulnerability prioritization metrics, organized into key categories: severity, exploitability, contextual and environmental factors, predictive metrics, and aggregated system-level factors. Severity metrics, particularly those based on the Common Vulnerability Scoring System (CVSS), are the most prevalent, focusing on potential impacts on confidentiality, integrity, and availability. Exploitability metrics assess the ease of exploitation, while contextual metrics incorporate asset criticality and operational constraints. Predictive metrics, though less frequently used, offer insights into potential future exploitation, and aggregation metrics provide a holistic view by combining various dimensions of risk. 

Our analysis also extends to the methodologies and validation approaches employed in the reviewed studies. Despite the significant progress made, many challenges remain. Current approaches often rely on static models, which struggle to integrate real-time threat intelligence and fail to adapt to rapidly changing attack surfaces. Additionally, the lack of transparency in machine learning (ML) based methods complicates their practical implementation and reduces their interpretability for decision-makers. The future of vulnerability prioritization lies in the development of adaptive, context-aware metrics that leverage dynamic threat intelligence to enhance risk assessments. Furthermore, the shift towards scalable, automated solutions, driven by artificial intelligence (AI) and ML, is expected to play an increasingly important role in managing large volumes of vulnerabilities in complex environments. This trend is particularly evident in the industrial platforms we analyze.

The contributions of this paper are threefold:

\begin{itemize} 
\item \textbf{Comprehensive systematic literature review}: We review 82 papers on vulnerability prioritization metrics, offering a detailed assessment of the current state of the field and its evolution up to December 2024. 
\item \textbf{A novel taxonomy of vulnerability prioritization metrics}: We propose a taxonomy that organizes metrics into severity, exploitability, contextual and environmental factors, predictive metrics, and aggregated system-level considerations. 
\item \textbf{Identification of research gaps and future directions}: We highlight key challenges, such as the need for dynamic and real-time prioritization models, the integration of adversarial and context-sensitive approaches, and the development of explainable AI-based techniques, providing clear recommendations for future research. \end{itemize}

Section \ref{sec:Preliminaries} introduces key concepts and definitions, including the major risk metrics and methodologies to lay the groundwork for the paper. Section \ref{sec:Methodology} outlines the research methodology employed in this survey. Sections \ref{sec:Impact} to \ref{sec:Predictive} provides an in-depth analysis of vulnerability prioritization metrics, as well as the utilized methodologies and validation techniques used in the 82 reviewed papers. Section \ref{sec:Discussion} discusses emerging challenges and trends. Finally, Section \ref{sec:Conclusion} offers concluding remarks.

\section{Preliminaries} \label{sec:Preliminaries}

\subsection{Key Concepts in Vulnerability Prioritization}

In complex systems, ensuring robust security against vulnerabilities is critical for maintaining operational integrity and safeguarding sensitive information. This sub-section introduces a formalized representation that focuses on the foundational dimensions of vulnerability prioritization: impact, exploitability, and contextual factors. These dimensions are selected for their direct relevance to calculating risk scores and their established role in prioritization frameworks. Additional categories, such as predictive and aggregated metrics, are later introduced with examples and integrated into the taxonomy presented. By linking the formalism to the taxonomy, this framework provides a structured approach for understanding and evaluating prioritization strategies.

\subsubsection{System and Vulnerability Representation}

We represent a complex system as a set $S$, consisting of $N$ components denoted as $C_i$, where $i \in \{1, 2, \dots, N\}$. Each component $C_i$ is associated with a set of vulnerabilities, where $V_{ij}$ denotes vulnerability $j$ in component $C_i$. Formally, each component is defined as:
\vspace{-2mm}
\[
S = \{ C_1, C_2, \dots, C_N \}, \quad C_i = \{ V_{i1}, V_{i2}, \dots, V_{iM} \}.
\]

$M$ is the amount of vulnerabilities in component $C_i$.

\subsubsection{Risk Score Calculation}

The risk score $RS_i$ for each component $C_i$ is a function of three key metrics:
\vspace{-2mm}
\[
RS_i = \alpha \cdot I_i + \beta \cdot E_i + \gamma \cdot C_i.
\]

Impact $I_i$ reflects the potential damage or disruption the vulnerability could cause in the component $C_i$. It is computed as the sum of the impact scores $I_{ij}$ of each vulnerability $j$ in component $C_i$:
\vspace{-2mm}
   \[
   S_i = \sum_{j=1}^{M} SV_{ij}.
   \]
   
The exploitability score $E_i$ represents how likely a vulnerability is to be exploited. It combines the time-to-exploit $TE_{ij}$ and the availability of an exploit $EA_{ij}$ for each vulnerability $j$ in component $C_i$:
\vspace{-2mm}
   \[
   E_i = \sum_{j=1}^{M} (TE_{ij} \cdot EA_{ij}).
   \]
   
Contextual or environmental factor $C_i$ captures the criticality of the component and its dependencies within the system. It is defined as the sum of the criticality level $L_i$ and system dependence $D_i$:
\vspace{-2mm}
   \[
   C_i = L_i + D_i.
   \]

\subsubsection{Objective Function}

The objective of the vulnerability prioritization process is to minimize the total risk score across all system components:
\vspace{-2mm}
\[
\text{Minimize} \quad R_{\text{total}} = \sum_{i=1}^{N} RS_i.
\]

This ensures that the system prioritizes the most critical vulnerabilities while accounting for severity, exploitability, and environmental factors.

The optimization process is subject to two main constraints, namely resource and operational constraints, respectively. Resource constraints suggest that the total remediation time must not exceed the available time. Operational constraints mean that the downtime for critical components must be minimized, ensuring that the system remains operational.

\subsection{Metrics for Vulnerability Prioritization}

\begin{figure}[h]
\centering
\includegraphics[width=0.55\textwidth]{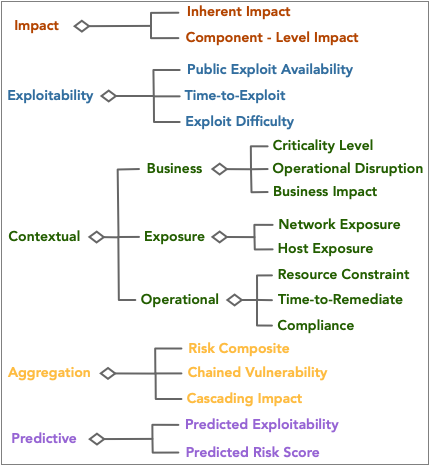}
\caption{Taxonomy of Vulnerability Prioritization.}
\label{fig:VulRankTax}
\end{figure}

In addition to the foundational metrics (i.e., impact ($I_i$), exploitability ($E_i$), contextual and environmental ($C_i$)) formalized earlier, this sub-section introduces a taxonomy that incorporates Predictive Metrics and Aggregated/System-Level Metrics, as presented in Fig. \ref{fig:VulRankTax}. This taxonomy is derived from a systematic review of 353 papers, of which 82 were selected for detailed analysis. By encompassing multiple dimensions of vulnerability prioritization, the taxonomy addresses the complexities of contemporary cybersecurity challenges, including regulatory pressures, the growing volume of vulnerabilities, and the need for holistic risk assessment frameworks.

The proposed taxonomy clarifies the role of each metric type and serves as a decision-making framework, enabling organizations to select metrics tailored to specific needs. For instance, impact and exploitability metrics are integral to technical assessments, while contextual metrics address operational relevance and compliance requirements. Predictive and aggregated metrics offer insights into future risks and system-wide dependencies, making them particularly valuable in resource-constrained or complex environments.

\textbf{Impact Metrics} quantify the inherent consequences of exploiting a vulnerability, focusing on its potential effects on system confidentiality, integrity, and availability. These metrics assess the intrinsic risk posed by a vulnerability, independent of contextual or exploitability considerations.

\textbf{Exploitability Metrics} evaluate the technical feasibility of exploiting a vulnerability, considering factors such as attack complexity, required privileges, user interaction, and exploit availability. These metrics prioritize vulnerabilities based on their ease of exploitation.

\textbf{Contextual and Environmental Metrics} incorporate deployment-specific and organizational factors to refine risk assessments. These metrics consider the criticality of affected components, business impact, operational constraints, and exposure levels, tailoring vulnerability prioritization to the system context.

\textbf{Predictive Metrics} provide forward-looking insights into the likelihood of future exploitation or evolving impacts. These metrics leverage statistical and machine learning models to anticipate potential threats and inform proactive risk management.

\textbf{Aggregated and System-Level Metrics} offer a holistic view of system-wide risk by integrating multiple dimensions into a comprehensive score. These metrics help identify vulnerabilities that contribute to cascading failures or affect interconnected components in complex environments.

\subsection{Methodologies for Vulnerability Prioritization}

This sub-section categorizes vulnerability prioritization methodologies into five main approaches: graph-based methods, ML and AI-based approaches, multi-objective optimization, rule-based and expert systems, and statistical methods.

\textbf{Graph Based Methods} model systems and vulnerabilities as interconnected nodes and edges, enabling the analysis of attack paths, dependencies, and cascading effects. These methods are particularly effective in complex environments where risks propagate across multiple assets. Structural models such as attack graphs, dependency graphs, and Bayesian networks help identify critical vulnerabilities and assess system-wide impacts, making them essential for prioritization in interconnected infrastructures.

\textbf{ML and AI based Approaches} leverage historical and real-time data to predict vulnerability risks and support data-driven decision-making. Techniques such as logistic regression, decision trees, and neural networks enable adaptive scoring, anomaly detection, and exploitation forecasting. These approaches are particularly advantageous in large-scale datasets, where traditional methods struggle to capture evolving attack patterns.

\textbf{Multi-Objective Optimization Methods} balance competing factors such as impact, exploitability, criticality, and resource constraints to generate optimized vulnerability rankings. By leveraging optimization algorithms, including genetic algorithms, integer programming, and evolutionary models, multi-objective approaches systematically evaluate and refine rankings to align with diverse security and operational goals.

\textbf{Rule-Based and Expert Systems} apply predefined rules, heuristics, and structured knowledge (e.g., ontologies, knowledge graphs) to assess vulnerabilities within specific domains. By integrating human expertise, they provide context-aware prioritization, ensuring reliable and repeatable assessments in environments where standardized models may be insufficient.

\textbf{Statistical Methods} use regression models, probabilistic techniques, and data distribution analysis to quantify risk factors and rank vulnerabilities. By identifying relationships between variables, these methods establish baseline risk metrics and refine prioritization rankings, providing a systematic, data-driven foundation for decision-making.

Table \ref{tab:methodologies_metrics} provides detailed mappings of the studied risk metric and utilized methodologies in the reviewed papers. 

\subsection{Existing Standards and Frameworks} \label{sec:ExistStandards}

The risk associated with vulnerabilities is often conceptualized using three factors, namely \textit{Probability}, \textit{Impact} and \textit{Threat} \cite{zeng2023illation}. \textit{Probability} quantifies the likelihood of exploitation \cite{elder2024survey}, \textit{Impact} assesses the consequences of a successful exploit, and \textit{Threat} identifies potential actors or circumstances exploiting the vulnerability.

CVSS \cite{cvss} is the leading vulnerability prioritization method. CVSS version 3 (V3), for instance, categorizes metrics into Base, Temporal, and Environmental groups. Base metrics include Exploitability (Attack Vector, Attack Complexity, Privileges Required, User Interaction), Scope, and Impact (Confidentiality, Integrity, Availability). Temporal metrics reflect dynamic aspects like exploit techniques and patch availability. Environmental metrics consider deployment contexts. These metrics are aggregated to generate a severity score for vulnerabilities.

%Then \textit{CVSS} v3 rates vulnerabilities as \textit{'none'} (0.0), \textit{'low'} (0.1-3.9), \textit{'medium'} (4.0-6.9), \textit{'high'} (7.0-8.9), or \textit{'critical'} (9.0-10.0).

The Exploit Prediction Scoring System (EPSS) \cite{epss} is a statistical model that estimates the likelihood of a vulnerability being exploited in the wild within the next 30 days \cite{jacobs2021exploit}. It provides a score between 0 and 1, where a higher score indicates a higher likelihood of exploitation. It uses logistic regression to evaluate features such as software vendor, exploit code availability, vulnerability characteristics, and associated references. 
%FIRST team provides EPSS API for easy access to exploit-probability scores.

In addition to CVSS and EPSS, the cybersecurity landscape employs various other standards and frameworks for vulnerability assessment and management. These include Common Platform Enumeration (CPE) \cite{cpe}, Common Weakness Enumeration (CWE) \cite{cwe}, Industrial Control Systems Cyber Emergency Response Team (ICS-CERT) \cite{ICSCERT} advisories. The Security Content Automation Protocol (SCAP) \cite{SCAP} provides a standardized approach to maintaining system security. Compliance frameworks such as North American Electric Reliability Corporation Critical Infrastructure Protection (NERC CIP) \cite{NERCCIP} and Payment Card Industry Data Security Standard (PCI DSS) \cite{PCIDSS} further guide organizations in maintaining robust cybersecurity practices. These diverse tools and standards collectively form a comprehensive ecosystem for vulnerability prioritization and risk management.

\section{Related Works}

Le et al. \cite{le2022survey} provide a comprehensive overview of data-driven software vulnerability (SV) assessment and prioritization, focusing on the use of ML, deep learning (DL), and NLP techniques to automate tasks in the SV management lifecycle. However, their scope is limited to the phases between SV discovery and remediation, excluding studies that rely solely on manual analysis or descriptive statistics.

Elder et al. \cite{elder2024survey} focus on methods for assessing the exploitability of vulnerabilities, categorizing them into manual CVSS-based assessments, automated deterministic assessments, and automated probabilistic assessments.

In contrast, our work introduces a broader taxonomy that includes compliance and contextual metrics, which have received limited attention in prior surveys. Furthermore, we analyze real-world challenges, such as explainability and vulnerability data quality, which extend beyond the data-driven focus of \cite{le2022survey}. By addressing these gaps, our study provides actionable insights into improving vulnerability prioritization frameworks for both research and industrial applications.

\section{Methodology} \label{sec:Methodology}

\subsection{Data Sources}

We conducted a systematic literature review using four major academic databases: ACM Digital Library, Scopus, IEEE Xplore, and Google Scholar (primarily for snowballing). To ensure comprehensive coverage, we formulated structured queries incorporating key terms related to vulnerability prioritization, risk assessment, exploitability, and cybersecurity frameworks. Papers published before December 2024 were queried, yielding 98 results from ACM Digital Library, 239 from Scopus, and 130 from IEEE Xplore. After deduplication and merging across databases, 353 unique papers remained.

\subsection{Selection Process}

We applied a two-stage filtering process. First, we conducted title and abstract screening, during which papers were excluded if they did not explicitly address vulnerability prioritization, risk-based decision-making, or security metric evaluation. We then continued with full-text review, whereby papers were assessed for methodological depth, use of structured risk metrics, and validation techniques. The following exclusion criteria were applied:

\begin{itemize}
    \item Irrelevant Content: Studies focusing solely on patch management, general cybersecurity frameworks, or qualitative discussions without prioritization-specific analysis.
\item Non-English Publications: To ensure interpretability and avoid translation inconsistencies.
\item Non-Peer-Reviewed Sources: Including white papers, blog posts, and non-academic industry reports.
\end{itemize}

Applying these criteria, 78 papers were selected for detailed review. We further conducted forward and backward snowballing, examining references in the selected papers and identifying additional citations. This process led to the inclusion of 4 additional studies, resulting in a final dataset of 82 papers for in-depth analysis.

\subsection{Data Extraction and Analysis}

The data extraction process was designed to systematically capture key aspects of each selected study to facilitate a thorough comparative analysis. Beyond collecting fundamental contextual information, such as the study’s objectives, research methodology, key findings, and evaluation or validation techniques, we performed a detailed manual extraction of specific, targeted data points relevant to vulnerability prioritization.

To enhance reliability, two independent reviewers annotated each paper, followed by consensus discussions to resolve discrepancies. In cases of ambiguity (e.g., distinguishing between rule-based and statistical models), a third reviewer conducted tie-breaking assessments to ensure objective categorization.

The extracted data points include:

\begin{itemize}
    \item We documented vulnerability prioritization metrics, focusing on severity, exploitability, contextual factors, predictive indicators, and aggregation methods, and other novel metrics to provide a comprehensive view.
    \item The risk assessment methodologies and frameworks employed in the studies were systematically cataloged. These included graph-based approaches, rule-based systems, ML techniques, multi-objective and statistical methods. We also examined whether the risk assessments were static or dynamic in nature.
    \item Validation methods, such as case studies, controlled experiments, simulations, and interviews, were analyzed to evaluate reliability and applicability across contexts.
    \item Studies were assessed for their use of real-world data (e.g., NVD, CVE, industry reports, threat intelligence feeds) versus synthetic datasets, highlighting the extent of real-world applicability.
    \item We examined alignment with security standards (e.g., ISO/IEC 27005, MITRE ATT\&CK), including custom adaptations for specific research objectives.
    \item We documented the challenges and limitations reported in each study, such as issues with scalability, difficulties in integrating new data sources, reliance on static metrics, lack of adaptability to real-time threats, and concerns regarding the explainability of advanced models, particularly those based on ML techniques. 
\end{itemize}

This comprehensive extraction process lays the foundation for the development of a novel taxonomy that categorizes vulnerability prioritization metrics and methods, providing a structured understanding of the field's current state and future directions. Such analysis also enables a deeper understanding of the methodologies and frameworks used in the field, providing a foundation for comparative analysis and the identification of trends, gaps, and emerging practices in vulnerability  prioritization.

\section{Impact Metrics}

\label{sec:Impact}

\subsection{Definition and Importance}

\textbf{Impact Metrics} quantify the inherent consequences of a vulnerability's exploitation, focusing on its effects on the confidentiality, integrity, and availability (CIA triad) of the affected system. These metrics evaluate the intrinsic risk posed by a vulnerability without considering contextual or exploitability factors.

\textbf{Inherent Impact Score ($IIS_{ij}$)} represents the inherent impact of vulnerability $j$ in component $i$, often derived from standardized frameworks such as the CVSS base score. Over 70\% of reviewed studies, including \cite{walkowski2021vulnerability, kia2024cyber, angelelli2024robust}, have utilized CVSS as the primary metric for vulnerability prioritization. However, CVSS base score has limitations, including its static nature \cite{spring2021time, le2021survey} and lack of contextualization. While widely adopted, CVSS base score does not reflect evolving threats such as active exploitation or new attack vectors. Additionally, it overlooks operational context; a high CVSS base score may pose minimal risk in a well-segmented environment, while a low score could endanger critical infrastructure in less secure systems.

%Recent research aims to enhance static vulnerability assessments by incorporating dynamic models and real-time exploitability predictions. For instance, EPSS \cite{yoon2023vulnerability, jacobs2023enhancing} addresses the limitations of static scoring systems by introducing predictive components. More details on this will be provided in the next sub-section. 

\textbf{Component-Level Impact ($CLI_i$)} measures the potential disruption caused by a vulnerability in component $i$ to system performance, data security, or critical functionality \cite{nourin2021measuring}. 

\subsection{Methodology Trends}

Impact metrics are most frequently used in rule-based and expert systems (21 studies), graph-based approaches (17 studies), and machine learning (17 studies). Multi-objective methods (8 studies) are used to a lesser extent, while statistical methods appear in only 6 studies. Below, we discuss key methodologies and representative studies.

Rule-Based and expert systems rely on predefined rules to assess vulnerability impact. For example, \cite{kurniawan2023automation} presents a rule-based approach for automating the quantification of security risks related to SQL injection and cross-site scripting attacks using CVSS. By leveraging vulnerability reports from NVD and real-time dynamic analysis of attack vectors (e.g., attacker’s IP location, privilege level, and network proximity), the system calculates CVSS vectors to assign risk scores to detected attacks.  

\cite{dissanayaka2020vulnerability} adjusts vulnerability severity scores by modifying the exploitability metrics of the CVSS framework, specifically Attack Vector (AV) and Attack Complexity (AC), based on the target environment's topology and implemented security mechanisms. This adjustment process results in a dynamic vulnerability score that reflects the exploitability of a vulnerability within a specific ICS environment, moving beyond generic CVSS scores to provide a more context-aware risk assessment. This paper also considers the impact on various system components (e.g., host, network, application, and containers) and uses root cause analysis to identify underlying factors contributing to vulnerabilities.

Graph-based approaches model vulnerabilities as nodes in a network, incorporating attack paths and dependencies. Attack tree \cite{li2023security} and Bayesian network \cite{wang2022automotive, wang2020bayesian, chatzipoulidis2015information} based approaches enhance vulnerability impact estimation. Specifically, vulnerability risk assessment using directed graph models necessitates quantifying nodes and paths with metrics categorized into node-, path-, and probabilistic-metrics. For example, \cite{li2023security} constructs an attack tree model that maps out potential attack paths, where the root node represents the ultimate attack goal, intermediate nodes denote steps in the attack sequence, and leaf nodes signify specific attack methods. Logical relationships between nodes (e.g., AND, OR and SAND) are used to model different attack scenarios. CVSS is applied to assess the severity of vulnerabilities at each leaf node, while a subjective-objective weighting method combines expert judgment with objective data to calculate the occurrence probability of security events. 

ML approaches are utilized to automate vulnerability scoring using text mining and ML techniques  \cite{aivatoglou2022rakel, costa2022predicting, hoque2023risk, mehta2023effect, pan2024towards}. For example, \cite{nourin2021measuring} recalculates CWE base scores using context similarity, a metric that measures the semantic similarity between CWE and CVE descriptions. Additionally, the paper uses weighted CWE frequency, where the impact of each weakness is adjusted based on how often it appears in the software. These metrics combine to generate an overall software security score.

Multi-objective methods strive to balance multiple conflicting criteria in vulnerability prioritization, considering factors beyond just impact. A more detailed discussion can be found in later sections where impact is typically treated as only one of several prioritization criteria.

Statistical approaches focus on empirical validation and predictive modeling of security impact. For instance, \cite{angelelli2024robust} propose a statistical framework using mid-quantile regression to prioritize vulnerabilities and introduce agreement of grounded rankings as accuracy measure, which maintains rank invariance despite incomplete information. This framework incorporates ordinal and quantitative metrics, including CIA, access vector and complexity from the NVD, alongside quantitative data like the number of vulnerable hosts and exploit availability.

\cite{holm2012empirical} employs statistical methods to analyze the effectiveness of various system-level vulnerability metrics by examining their correlation with the time-to-compromise metric during actual cyber-attacks. The study uses Pearson correlation to assess how well different CVSS-based metrics, including weakest link models and aggregated vulnerability scores, predict system security. Additionally, it explores the concept of vulnerability exposure, measuring the duration that vulnerabilities remain unpatched, and how this impacts system risk.

\begin{tcolorbox}[colback=gray!10!white, colframe=blue!75!black, left=2mm, right=0mm, top=1mm, bottom=1mm, boxrule=0pt, sharp corners, before skip=10pt, after skip=10pt, coltitle=black] 
\textbf{Observation:} CVSS remains the dominant standard for vulnerability prioritization despite its static nature and lack of contextualization. Hybrid models, integrating CVSS with Bayesian networks, attack graphs, or ML-based ranking, show promise in refining impact assessments by incorporating dynamic system conditions and adversary behaviors.
\end{tcolorbox}

\section{Exploitability Metrics}

\label{sec:Exploitability}

\subsection{Definition and Importance}

\textbf{Exploitability Metrics} assess the technical ease of exploiting a vulnerability, focusing on factors such as attack complexity, required privileges, user interaction, and the availability of public exploit tools or code. These metrics operate independently of the deployment environment or contextual considerations.

\textbf{Public Exploit Availability ($PEA_i$)} indicates whether a publicly known exploit exists for vulnerabilities in component $i$. A known exploit increases urgency for remediation \cite{samtani2022linking, seker2023xvrs}.

\textbf{Time-to-Exploit ($TTE_i$)} estimates how soon a functional exploit is likely to emerge in the wild for vulnerabilities in component $i$. This metric is essential for prioritizing vulnerabilities that may soon have active exploits but are not currently being used in attacks \cite{costa2022challenges}.

\textbf{Exploit Difficulty ($ED_i$)} measures the difficulty of exploiting a vulnerability, considering factors like user interaction and the presence of security mechanisms such as firewalls or sandboxing. Exploit difficulty can be adjusted based on environmental factors, such as the presence of firewalls or intrusion detection systems \cite{yoon2023vulnerability}.

\begin{tcolorbox}[colback=gray!10!white, colframe=blue!75!black, left=2mm, right=0mm, top=1mm, bottom=1mm, boxrule=0pt, sharp corners, before skip=10pt, after skip=10pt, coltitle=black]
\textbf{Observation:} Exploitability metrics have evolved beyond static scoring systems to integrate signals from exploit repositories, social media, and dark web discussions. ML and graph-based approaches are increasingly used to predict exploitability, but challenges remain in leveraging noisy, adversarial, and incomplete threat intelligence data.

\textbf{Open Problems:}  

1. Addressing potential data poisoning and misinformation in social media and dark web sources.  

2. Integrating threat intelligence from diverse sources (e.g., CVEs, MITRE ATT\&CK, dark web, malware databases) into unified exploitability risk frameworks.  
\end{tcolorbox}

\subsection{Methodology Trends} 

Exploitability metrics are predominantly used in ML-based methods (17 studies) and graph-based approaches (13 studies), with rule-based systems (9 studies) and statistical methods (6 studies) playing a secondary role. Multi-objective (2 studies) is the least utilized.

ML techniques are widely applied to predict exploitability by analyzing textual and structured vulnerability data. For example, \cite{costa2022challenges} introduces the vulnerabilities’ risk of exploitation system (V-REx), which prioritizes software security patches based on the estimated probability of exploitation. This probability is computed using interconnected neural networks optimized with a genetic algorithm. V-REx incorporates multiple factors, including exploit availability, severity, and the likelihood of exploitation, to rank vulnerabilities effectively.

\cite{iannone2024early} employs multiple ML classifiers to predict the exploitability of newly disclosed vulnerabilities using CVE descriptions, CVSS base scores, and online discussions from Security Focus and Exploit Database. They evaluate multiple classifiers, finding Logistic Regression to provide the best trade-off between precision and recall, while Random Forest achieves the highest precision. Pre-trained Large Language Models (LLMs) underperform, acting as majority-class predictors. Their proposed method also integrates SMOTE (or synthetic minority over-sampling technique) for data balancing, which improves prediction accuracy in imbalanced datasets. 

\cite{samtani2022linking} uses a combination of a bidirectional long-short term memory (Bi-LSTM) network and attention mechanisms to link exploits from Dark Web hacker forums to known vulnerabilities. Based on the generated exploit-vulnerability linkages, this metric incorporates factors like exploit post date, the number of vulnerabilities on a device and the age of associated exploits to create an aggregated risk score for devices.

\cite{seker2023xvrs} integrates the availability of exploits, directly influencing the risk score by indicating immediate exploitation potential. In addition, the authors consider the volume of social media activity (e.g., tweets, likes, and retweets), which reflects public attention; engagement in public code repositories (e.g., the number of forks and stars on GitHub), indicating the level of exploit or mitigation development; and mentions in cybersecurity news and advisories, which highlight the vulnerability's prominence in the industry. They also consider dark web activity as a risk indicator, with mentions on hacker forums suggesting active interest in exploitation. 

Graph-based methods model exploitability by capturing relationships among vulnerabilities, assets, and attacker behaviors. \cite{chen2019using} introduces the CVE-Author-Tweet (CAT) graph, a multi-layer directed graph that predicts vulnerability exploitability using Twitter discussions, independent of CVSS scores. It comprises three interconnected layers: CVE graph, linking vulnerabilities via co-mentions in tweets; Author graph, modeling user interactions (e.g., mentions, follower-followee); and Tweet graph, representing tweet relationships (e.g., retweets, shared CVEs). Cross-layer edges propagate information across authors, tweets, and vulnerabilities, enabling real-time tracking of vulnerability discussion dynamics.

Rule-based methods rely on predefined criteria to assess exploitability likelihood, while statistical approaches analyze historical data to infer exploitability trends. For instance, \cite{falco2018iiot} prioritizes SCADA vulnerabilities by analyzing how common certain vulnerability types are (CWE density), how many have publicly available exploits (CWE exploit density), and the potential impact and exploitability of each vulnerability based on their CVSS scores.

\cite{yoon2023vulnerability} incorporates statistical modeling to estimate the probability of successful exploitation based on exploit code availability and exploit use probability. The system integrates data from various sources, including the MITRE ATT\&CK framework and CVE databases, and then employs a structured mapping methodology to connect CVEs to ATT\&CK-defined adversary tactics, techniques, and procedures (TTPs) used by attackers. Their method is tailored for vulnerability assessment in Operational Technology (OT) and ICS environments, with a focus on Industrial Internet of Things (IIoT) devices.

\section{Contextual and Environmental Metrics}

\label{sec:Contextual}

\textbf{Contextual and Environmental Metrics} evaluate vulnerability risk based on system-specific and organizational factors, such as the criticality of the affected component, potential disruption to business operations, and broader financial or reputational implications. These metrics adapt risk assessments to reflect the unique context of the deployment environment.

\subsection{Definition and Importance} 

\subsubsection{Business and System Impact Metrics} 

These metrics measure how the vulnerability's impact extends beyond the technical scope to affect broader business and system objectives.

\textbf{Criticality Level ($CL_i$)} reflects the importance of component $i$ within the overall system. Components critical to business operations or system functionality receive a higher priority in remediation decisions \cite{ahmadi2022automated, hu2023cost}.

\textbf{Operational Disruption Risk ($OD_i$)} measures the potential for a vulnerability to disrupt essential business processes or cause system downtime \cite{longueira2022novel, longueira2022gotta, ghosh2015netsecuritas, keskin2021scoring}.

\textbf{Business Impact ($BI_i$)} evaluates the potential financial, regulatory, or reputational damage from exploiting a vulnerability in component $i$ \cite{yadav2022smartpatch, hore2022towards}.

%User exposure ($UE_i$) reflects the number of users or customers that could be affected by a vulnerability. Vulnerabilities in components with high user exposure are prioritized for remediation to minimize disruption.

\subsubsection{Network and Host Exposure}

These metrics measure how accessible vulnerable components are from external networks or internally exposed surfaces. These metrics help prioritize vulnerabilities based on their level of exposure to potential attackers.

\textbf{Network Exposure level ($NE_i$)} evaluates the degree of exposure of component $i$ to external networks, such as the internet or unsecured network segments. Vulnerabilities in externally facing components, such as web servers or public APIs, pose higher risks and are prioritized accordingly \cite{mahmood2021prioritizing, cheng2023network}.

\textbf{Host Exposure Level ($HE_i$)} measures the exposure of component $i$ to potential internal threats due to host-level mis-configurations, open ports, or unnecessary services \cite{wang2020vulnerability, zeng2023illation}.

\subsubsection{Operational Feasibility Considerations}

These metrics account for factors such as time-to-remediation, resource constraints, and regulatory requirements. These metrics are crucial for understanding the practical challenges in vulnerability remediation and ensuring compliance with industry standards.

\textbf{Time-to-Remediation ($TTR_i$)} estimates the time required to apply a patch or mitigation for component $i$. Vulnerabilities with shorter remediation timelines are prioritized to minimize the window of exposure, while those with longer timelines may necessitate interim measures, such as temporary mitigations or enhanced monitoring \cite{walkowski2021automatic, walkowski2021vulnerability, reyes2022environment, farris2018vulcon}.

\textbf{Resource Constraints ($RC_i$)} reflects the availability of resources (e.g., personnel, budget, and tools) needed to remediate vulnerabilities in component $i$. Limited resources necessitate prioritizing vulnerabilities that can be resolved with minimal disruption or cost, balancing overall risk mitigation with operational efficiency \cite{farris2018vulcon}.

\textbf{Compliance and Regulatory Impact ($CR_i$)} assess the legal and regulatory obligations associated with component $i$. Vulnerabilities in components subject to regulatory frameworks (e.g., GDPR and HIPAA) are prioritized for timely remediation to avoid penalties, legal action, or reputational damage \cite{farris2018vulcon}. 

Compliance regimes, such as PCI DSS, HIPAA, or NERC CIP, impose specific timelines (e.g., service level objectives (SLOs) \cite{qiu2020firm} and service level agreements (SLAs) \cite{luna2015quantitative}) for vulnerability remediation. These requirements often prioritize regulatory adherence over traditional security risk metrics, shaping how vulnerabilities are addressed in practice. For example, vulnerabilities with medium CVSS scores may be prioritized over high-severity ones if they affect systems covered under compliance mandates with tight remediation deadlines. 

%PCI DSS compliance is particularly relevant to the financial and enterprise sectors.

\cite{farris2018vulcon} highlights a critical challenge in cyber-insurance: establishing standardized metrics to quantify organizational cybersecurity risk. This article emphasizes the need to scientifically identify key vulnerability features that most significantly impact overall threat exposure, particularly in the context of regulatory frameworks like HIPAA and PCI DSS.

\cite{dissanayaka2020vulnerability} suggests securing data in a MongoDB-based system to comply with HIPAA regulations, especially concerning authentication, authorization, encryption, and auditing. The specific use of HIPAA as a metric for vulnerability prioritization does not feature in the paper’s primary methodology though.

\begin{tcolorbox}[colback=gray!10!white, colframe=blue!75!black, left=2mm, right=0mm, top=1mm, bottom=1mm, boxrule=0pt, sharp corners, before skip=10pt, after skip=10pt, coltitle=black] 
\textbf{Observation:} Regulatory frameworks, such as SLAs and compliance mandates, often dictate remediation priorities, sometimes overriding risk-based assessments. However, existing models struggle to integrate these constraints effectively, leading to misaligned prioritization.

\textbf{Open Problems:}  

1. Incorporating compliance requirements into contextual risk models without overshadowing technical risk factors.

2. Developing frameworks that adjust to evolving regulatory landscapes while maintaining security efficiency.

3. Balancing risk-based, operational, and regulatory constraints within a unified decision-making framework.
\end{tcolorbox}

\subsection{Methodology Trends}

Contextual metrics are heavily utilized in graph-based approaches (21 studies), emphasizing their ability to model system dependencies and environmental factors. Rule-based systems (17 studies) are also widely applied, integrating contextual information into predefined scoring mechanisms. ML-based approaches (8 studies) leverage structured data to refine prioritization models. Multi-objective and statistical approaches appear less frequently (5 and 3 studies, respectively).

Graph-based models provide a structured approach to vulnerability assessment by capturing attack sequences, exploit dependencies, system relationships, and contextual risk factors. Various studies leverage different graph structures to enhance vulnerability prioritization and risk estimation. 

Attack graphs represent sequences of attack steps that an adversary can take to compromise a system \cite{yan2022cyber, angelini2018vulnus, garg2018empirical, ouedraogo2013towards}. Nodes correspond to vulnerabilities, while edges depict potential exploit transitions. These graphs facilitate risk quantification, exploit dependency analysis, and multi-step attack simulations.

Exploit dependency graphs is a specialized subset of attack graphs, focusing on exploit dependencies and illustrating how one exploit enables another. For example, \cite{ghosh2015netsecuritas} introduces an in-degree dependency metric, measuring the number of attack paths in which an exploit appears. Exploits with higher in-degrees are deemed more critical due to their involvement in multiple attack scenarios. 

Heterogeneous information network (HIN) extend attack graph methodologies by modeling multiple types of entities and relationships, such as hosts, vulnerabilities, and access controls. Unlike traditional attack graphs, HINs provide a more scalable and adaptable framework for network-aware vulnerability prioritization. \cite{wang2020vulnerability} introduces a HIN-based risk assessment model, representing vulnerabilities, hosts, and their interconnections. The model integrates CVSS impact and exploitability metrics alongside contextual factors such as network topology, component importance, and exposure levels. A PageRank-inspired ranking algorithm iteratively refines risk scores, prioritizing vulnerabilities based on their effect on network assets. The model’s extensibility allows for the incorporation of additional contextual factors, improving the adaptability of vulnerability rankings.

System dependency graphs model software/hardware dependencies and how vulnerabilities propagate across interconnected components. For instance, \cite{longueira2022novel} proposes an extended dependency graph (EDG) to model industrial system vulnerabilities, incorporating CVSS-based risk metrics. The EDG tracks vulnerability origins, dependency chains, and patch prioritization, validated through a case study on OpenPLC. Building on this work, \cite{longueira2022gotta} extends this work by aggregating CVSS scores, incorporating functionality disruption and deployment context into vulnerability rankings.

A complementary approach integrates graph-based patch prioritization with game theory (GT). PatchRank \cite{yadav2019patchrank} applies GT to model the interaction between attackers and defenders, where attackers maximize exploitation, and defenders seek optimal patching strategies. PatchRank further employs a two-player non-cooperative game to calculate the probability of attacks and defenses, with a mixed Nash equilibrium used to determine the optimal patching strategy. The model computes risk scores iteratively, considering network topology, asset criticality, and resource constraints. SmartPatch \cite{yadav2022smartpatch} builds on PatchRank, introducing a residual impact score to refine prioritization based on patch effectiveness over time.

Vulnerability relation networks focus exclusively on relationships between vulnerabilities, identifying structural risk factors through graph-theoretic metrics. \cite{mahmood2021prioritizing} employs network-level and node-level metrics to rank vulnerabilities based on their structural significance. Network-level metrics such as average degree, network density, and clustering coefficient assess how vulnerabilities are interconnected within a network. Node-level metrics like betweenness centrality and closeness centrality evaluate the influence and position of vulnerabilities within the network. These metrics reveal previously overlooked vulnerabilities, showing that some low-CVSS vulnerabilities can still be critical due to their centrality in attack propagation.

Knowledge graphs (KGs) integrate multi-source intelligence, providing semantic contextualization of vulnerabilities and their relationships.. In industrial and network environments, these models facilitate vulnerability ranking by integrating path-based reasoning and contextual relationships. For instance, \cite{wang2023critical} leverages critical path aggregation that uses a KG to identify exploitable vulnerabilities and evaluate and weight multiple relation paths between an attacker and vulnerabilities, based on connectivity, privilege levels, and attacker accessibility. A query relation reasoning mechanism evaluates attack feasibility based on system constraints, prioritizing vulnerabilities that enable high-impact adversary objectives.

Similarly, \cite{cheng2023network} applies attack path analysis through KGs, mapping vulnerabilities to assets and attack routes.
A modified PageRank algorithm ranks vulnerabilities based on severity, exposure scope, and asset criticality.

\begin{tcolorbox}[colback=gray!10!white, colframe=blue!75!black, left=2mm, right=0mm, top=1mm, bottom=1mm, boxrule=0pt, sharp corners, before skip=10pt, after skip=10pt, coltitle=black]
\textbf{Open Problems:}  

1. Enhancing attack and knowledge graphs for large-scale systems while maintaining computational efficiency.  

2. Establishing unified graph-based vulnerability assessment frameworks for consistent cross-model integration.  

3. Developing real-time updates for evolving threat landscapes, ensuring resilience against adaptive attacks.  
\end{tcolorbox}

Rule-based approaches rely on structured heuristics to prioritize vulnerabilities based on contextual factors. CAVSS \cite{jung2022cavp} employs expert-validated rules to compute temporal vulnerability scores, integrating base, temporal, and environmental factors into a comprehensive risk metric. It considers exploit maturity, remediation levels, and report confidence, assigning higher scores to vulnerabilities with verified exploits or official fixes. The system incorporates real-world contextual data from vendors and authoritative sources to tailor prioritization to specific environments.

\cite{mehri2023automated} introduces a three-phase automated risk management system containing predefined steps such as patch testing, verification, and rollback. A feedback loop refines prioritization based on historical patch success rates and organizational constraints, such as patching time, expert availability, system downtime, and software dependencies. Similarly, \cite{ahmadi2022automated} filters vulnerabilities using predefined rules based on asset inventories and vulnerability scans, applying access policies to classify application/unit/services (AUS) as internal or external. Expert interviews further refine patch prioritization criteria and scoring.

\cite{keskin2021scoring} enhances prioritization by modeling functional dependencies between assets and business processes. The system evaluates how operability loss in one asset propagates to dependent assets and ultimately impacts critical business functions. Meanwhile, \cite{walkowski2021automatic} extends CVSS by incorporating environmental metrics (e.g., Collateral Damage Potential, Target Distribution) alongside time-to-remediation and asset value, enabling dynamic vulnerability assessment. This approach integrates real-time data from vulnerability scans and asset management tools to optimize patch prioritization. On top of this, \cite{walkowski2021vulnerability} further automates the CVSS environmental scoring process.

ML models automate the integration of contextual factors for vulnerability prioritization. \cite{hore2022towards} combines CVSS scores with network context (e.g., asset importance, existing defenses) in a decision-support system that first ranks vulnerabilities for mitigation and then optimally assigns tasks to security personnel based on their skillsets.

Multi-objective approaches balance risk prioritization under multiple constraints. \cite{colombelli2022multi} develops evolutionary algorithms to generate ranked vulnerabilities, optimizing factors such as risk, vulnerability age, and application importance. A fitness function evaluates solutions, refining rankings iteratively. They also integrate a multi-objective post-optimization process that fine-tunes the solutions by adjusting the ranking to correct any inconsistencies. Additionally, SmartPatch \cite{yadav2022smartpatch}, a framework for SCADA system patch prioritization, calculates a Residual Impact Score (RIS) to assess unpatched vulnerabilities' effects while integrating functional and topological dependency scores to measure subsystem criticality. Exploitability and patch interdependencies are incorporated to optimize patching, with Nash equilibrium-based game theory determining optimal strategies under resource constraints.

Statistical approaches provide quantitative models for contextual vulnerability risk. \cite{reyes2022environment} extends CVSS with risk factor (RF) metrics, incorporating probability of occurrence and impact assessments. The model calculates total vulnerabilities per IP (TV) to evaluate system-wide exposure and average organizational risk (AOR) to adjust CVSS scores based on environment-specific factors. Additionally, probability of exploitation (PoE) and average remediation time (AVT) guide prioritization, emphasizing vulnerabilities with high likelihood of exploitation and long mitigation times.

\section{Aggregated and System-Level Metrics}

\label{sec:Aggregation}

\subsection{Definition and Importance}

\textbf{Aggregated and System-Level Metrics} provide a holistic view of system risk by combining multiple dimensions of risk into a single, comprehensive score. This category accounts for the overall impact of vulnerabilities on the entire system, including attack paths and system-wide dependencies. These metrics are particularly useful for prioritization in environments with complex dependencies, such as supply chains and cloud systems.

\textbf{Composite Risk Score ($CR_i$)} for component $i$ combines metrics such as impact, exploitability, and exposure into a single score, often weighted based on organizational priorities \cite{zeng2023illation, zeng2021licality, bulbul2014cyber}. 

\textbf{Chained Vulnerability Impact ($CV_i$)} assesses the risk of multi-step attack paths where vulnerabilities in one component enable lateral movement or privilege escalation across the system \cite{sharma2023analysis}.

\textbf{Cascading Impact ($CI_i$)} measures the potential for cascading failures across interconnected components due to a single vulnerability. High dependency on vulnerable components can elevate system-wide risk \cite{cheimonidis2024dynamic, khaledian2018power, yan2022cyber}.

\begin{tcolorbox}[colback=gray!10!white, colframe=blue!75!black, left=2mm, right=0mm, top=1mm, bottom=1mm, boxrule=0pt, sharp corners, before skip=10pt, after skip=10pt, coltitle=black] 
\textbf{Observation:} Aggregated metrics offer a holistic risk assessment by integrating exploitability, impact, and system dependencies. However, averaging and weighting mechanisms can obscure critical vulnerabilities, reducing interpretability and prioritization accuracy.

\textbf{Open Problems:} 

1. Dynamically weighting risks based on asset importance, evolving attack paths and real-time dependencies.

2. Capturing how local vulnerabilities escalate into system-wide threats.

3. Enhancing transparency to prevent information loss in aggregated scores.

\end{tcolorbox}

\subsection{Methodology Trends}

Research on vulnerability prioritization utilizing aggregation metrics spans rule-based (10 studies), graph-based (7), ML-based (3), and multi-objective optimization (3), and statistical (1) approaches.

Rule-based methods aggregate multiple risk factors by applying structured heuristics for vulnerability prioritization. \cite{almazrouei2022internet} employs an attack graph-based penetration testing model to aggregate risk across attack paths in IoT networks. The model integrates exploitability metrics (e.g., programming language, exploit availability), impact metrics (e.g., CVSS scores, privilege escalation potential), and contextual factors (e.g., network topology, vendor reputation, exploit age). Adjacency matrices and NetworkX tools enable the aggregation of attack paths for critical path analysis, identifying shortest and most impactful attack routes.

\cite{hu2023cost} aggregates urgency scores by combining CVSS base scores, exploit maturity, and cloud-native security features. A cloud-native remediation level assesses existing security controls (e.g., firewalls, endpoint detection) to determine whether vulnerabilities require immediate patching or alternative mitigation. The risk reduction rate measures the effectiveness of mitigations, while an asset value score further prioritizes vulnerabilities affecting mission-critical infrastructure.

In power grid security, \cite{bulbul2014cyber} introduces an approach where the impact factor and vulnerability index evaluate cascading failures across substations. The intrusion credibility index assesses cyber intrusion risks based on external network access, while a priority ranking system categorizes substations by criticality and cascading effects.

Graph-based models provide network-aware risk assessments that incorporate attack path aggregation and asset interdependencies. For example, \cite{cheimonidis2024dynamic} introduces a model that dynamically adjusts vulnerability severity within ICS environments by integrating exploitability metrics which are then modified based on network topology and security mechanisms. The model uses attack trees to assess inter-dependencies between vulnerabilities and calculate aggregated scores for assets with multiple vulnerabilities. Also, fuzzy cognitive maps (FCM) simulate attack paths and provide a dynamic vulnerability score by integrating exploitability metrics and asset relationships.

For power grids, \cite{khaledian2018power} integrates fault chain theory with cascading failure analysis to assess vulnerabilities that contribute to system-wide failures. A vulnerability index is derived by combining power flow disruptions and fault propagation likelihood, providing a comprehensive risk metric for critical transmission lines.

\cite{yan2022cyber} develops a network security model for SCADA systems, calculating exploit probabilities across direct and indirect attack paths while considering factors like defense bypass and high-risk vulnerabilities. Attacker capabilities and defense mechanisms are incorporated into an exploitation likelihood model, while physical disruption metrics (e.g., minimum shedding load) assess aggregate system impact from cyberattacks.

\cite{sharma2023analysis} employs vulnerability tree analysis to evaluate the interconnectedness of vulnerabilities based on asset value, exploitation likelihood, and potential impact. This method enhances risk visualization by identifying critical paths where vulnerabilities propagate across interconnected systems.

Hybrid approaches aggregate ML-driven risk predictions with rule-based logical models to improve prioritization accuracy. For instance, the AVIA \cite{tatarinova2018avia} framework aggregates cross-dependency analysis and binary impact metrics to model system-wide vulnerability propagation. By mapping how vulnerabilities propagate across shared binaries and system objects, the framework generates risk scores that reflect cumulative exploitability effects.

\cite{zeng2023illation} integrates ML and rule-based logical programming to enhance vulnerability prioritization. The methodology employs neural networks and neuro-symbolic computing to learn adversary behavior patterns, vulnerability interactions, and network constraints. Probabilistic logic programming (ProbLog) refines risk assessments by incorporating host reachability and attack likelihood. Building upon LICALITY \cite{zeng2021licality}, \cite{zeng2023illation} introduces network-specific interaction constraints, considering vulnerable services, network topology, and adversary behavior. 

Multi-objective optimization methods aggregate vulnerability prioritization under competing constraints. \cite{farris2018vulcon} proposes VULCON that uses a mixed-integer multi-objective optimization algorithm to prioritize vulnerabilities for patching. VULCON optimizes conflicting objectives using goal programming and mixed-integer programming. Total vulnerability exposure (TVE) measures the overall exposure of unmitigated vulnerabilities within a network, while Time-to-vulnerability remediation (TVR) tracks the time between vulnerability discovery and remediation. The mitigation utility (MU) score prioritizes vulnerabilities based on severity, persistence, age, and the mission-criticality of the affected hosts. This method emphasizes the importance of mitigation utility scoring for vulnerabilities, based on their severity, persistence, and the mission-criticality of the assets they affect. 

Building on VULCON, \cite{shah2022vulnerability} introduces individual vs. multiple attribute value optimization, where individual attribute optimization ranks vulnerabilities based on a single criterion (e.g., persistence), while multi-attribute optimization refines prioritization using an aggregated function.

\section{Predictive Metrics}

\label{sec:Predictive}

\subsection{Definition and Importance}

\textbf{Predictive Metrics} evaluate how vulnerability risk evolves over time, using models to forecast future exploitation or impact to inform timely remediation. Predictive metrics forecast future risks by leveraging historical data, threat intelligence, and machine learning. These metrics are particularly valuable for proactive risk management, enabling organizations to prioritize vulnerabilities likely to be exploited in the near term. Their effectiveness relies heavily on high-quality, real-time data, which remains a persistent challenge.

\textbf{Predicted Time-to-Exploit ($PTTE_i$)} is a model-based estimation of how quickly a vulnerability in component $i$ might be exploited after discovery, based on historical trends and threat intelligence \cite{iannone2024early, yin2023empowering, figueiredo2022exploitability}.

Among existing models, EPSS \cite{jacobs2023enhancing, yoon2023vulnerability} stands out for its comprehensive feature set, incorporating over 1,400 attributes such as CVSS metrics, public exploit code availability, and social media mentions. EPSS is particularly effective for short-term exploit predictions (e.g., 30 days) due to its focus on dynamic threat signals.

\textbf{Predictive Risk Score ($PR_i$)} combines likelihood and impact forecasts into a single risk score generated by predictive models (e.g., ML techniques), offering a holistic view of future risks \cite{kia2024cyber, zeng2021licality}.

\subsection{Methodology Trends}

Predictive metrics primarily appear in ML-based methods (13 studies), followed by graph-based models (3 studies). Rule-based systems and statistical models show minimal adoption, with only 2 studies each incorporating predictive metrics.

ML models utilize historical and real-time threat intelligence to predict future vulnerability exploitation. These approaches integrate text analysis, latent risk estimation, and (un)supervised learning algorithms to refine predictive accuracy. For example, \cite{kia2024cyber} introduces a topic extraction-based risk prediction model, assigning cybersecurity topics to CVEs using Wikipedia-derived text features. A time-series risk score is computed by multiplying occurrence and impact metrics, and random forest models are trained on historical data to forecast future risks.

\cite{zhang2020dynamic} employs deep learning for time-dependent exploitability prediction, integrating neural networks to dynamically adjust risk scores. The model is trained on CVSS metrics, CWE categories, software attributes, and text-based features (TF-IDF bi-grams) to classify vulnerabilities exploitability. It incorporates two scheduling algorithms: a baseline NP-hard optimization method and a group-based scheduling method, which reduces computational complexity by dividing assets into groups.

\cite{figueiredo2022exploitability} presents V-REx, an neural network based vulnerability exploitability prediction system. It classifies vulnerabilities using three neural network models (standard, enhanced, interconnected enhanced), leveraging textual features, CVSS scores, and metadata from CVE/NVD sources. The system fine-tunes hyperparameters using an enhanced genetic algorithm.

Several studies integrate ML with optimization and rule-based reasoning to enhance predictive risk assessment. \cite{zeng2021licality} develops a neuro-symbolic model that prioritizes vulnerabilities based on past exploitability trends and future impact potential. A neural network analyzes historical threat data using latent semantic analysis (LSA), while criticality metrics (CVSS CIA impact, access complexity) refine prioritization. The model integrates these factors into a hybrid risk computation framework, combining ML-based probability estimation with rule-based reasoning.

\cite{hore2023deep} combines deep reinforcement learning (DRL) with integer programming to optimize resource allocation under uncertainty and limited security resources. The DRL agent anticipates critical vulnerabilities by integrating asset criticality, CVSS scores, and intrusion detection system (IDS) alerts, dynamically adjusting prioritization based on evolving threat conditions.

Graph-based methods predict exploitability by analyzing structural risk propagation and interdependencies within complex systems. \cite{yin2023empowering} improves exploitability prediction using graph-based ML. The model applies topological vulnerability graph analysis, incorporating PageRank, degree centrality, and label propagation to capture structural risk relationships. Additionally, heterogeneous graph neural networks (HGNNs) generate node embeddings, refining vulnerability risk prediction beyond traditional CVSS-based assessments.

\cite{chen2019vase} introduces a social media-enhanced vulnerability risk model, using graph convolutional networks (GCN) to analyze Twitter discussions for early vulnerability risk detection. Nodes represent CVEs, and edges capture semantic similarities in tweets. Attention-based embeddings refine feature extraction, enabling predictive CVSS scoring based on the first three days of Twitter discussions following a CVE’s public disclosure.

%\begin{tcolorbox}[colback=gray!10!white, colframe=blue!75!black, left=2mm, right=0mm, top=1mm, bottom=1mm, boxrule=0pt, sharp corners, before skip=10pt, after skip=10pt, coltitle=black]
%\textbf{Observation:} Predictive metrics refine vulnerability prioritization by forecasting exploitation likelihood and impact. While ML and graph-based models improve accuracy, challenges remain in real-time adaptation, explainability, and data quality.  

%\textbf{Open Problems:}  
%1. \textit{Real-Time Adaptation}: Enhancing predictive models to incorporate evolving threat signals with minimal delay.  
%2. \textit{Explainability}: Improving transparency in ML-based risk predictions for actionable insights.  
%3. \textit{Data Limitations}: Addressing data imbalance, scarcity, and bias in predictive models.  
%4. \textit{Graph-Based Forecasting}: Advancing adaptive graph models to capture evolving attack paths.  
%\end{tcolorbox}

\section{Discussions} \label{sec:Discussion}

Through the exploration of key metrics, methodologies, and validation techniques, this section synthesizes the current trends, challenges, and emerging directions in vulnerability prioritization research.

\subsection{Cross-Metric Trends and Insights}

\subsubsection{Metric Usage}

A quantitative analysis of the reviewed studies reveals the prevalence and gaps in metric usage. Impact metrics dominate, appearing in 60 of 82 studies, yet they often insufficient for capturing operational relevance or future risks. Exploitability metrics, featured in 40 studies, emphasize the technical feasibility of exploitation (e.g., network accessibility, exploit availability) but their integration with other dimensions remains limited. Contextual metrics, present in 48 studies, reflect a growing focus on asset criticality and operational relevance. Used in 20 studies, aggregated metrics combine multiple dimensions of risk but remain under-explored in system-wide prioritization frameworks. Predictive metrics, despite their potential for proactive decision-making, appear in only 17 studies, indicating a substantial opportunity for further research. Table \ref{tab:DataSourceSeverityMetrics} in the Appendix maps the reviewed studies, detailing their use of data sources and metric categories.

The integration of multiple metrics remains limited. Only 22 studies combine exploitability and contextual metrics, while predictive and aggregated metrics are seldom jointly applied (only in 2 studies). These findings emphasize the need for multidimensional approaches that incorporate operational and holistic risk assessments.

\subsubsection{Methodological Insights}

A balanced interest exists between data-driven techniques (graph-based, ML/AI) and expert-driven methods, each appearing in 24–27 studies. Graph-based methods are the most frequently employed, appearing in 27 of the reviewed papers. These methods leverage network or dependency graphs to model the relationships between system components or vulnerabilities \cite{wang2020vulnerability, longueira2022gotta, yadav2019patchrank, almazrouei2022internet, li2023security}. These graph-based approaches normally aim to propagate risks across interconnected nodes to identify critical vulnerabilities. ML and AI-based methods are utilized in 24 papers, focusing on predictive analysis and anomaly detection \cite{iannone2024early, hore2023deep, seker2023xvrs, costa2022challenges, fang2020fastembed}. These approaches rely on historical data to train models that prioritize vulnerabilities based on previous exploitation patterns or system configurations. Rule-based and expert systems are also present in 24 papers, employing predefined heuristics or expert knowledge to prioritize vulnerabilities \cite{kurniawan2023automation, jung2022cavp, wang2023critical, cheng2023network, walkowski2021vulnerability}. These systems, though less flexible than ML, are effective in environments with well-established risk metrics and clear vulnerability classifications. Multi-objective optimization are utilized in 10 papers, balancing different aspects such as severity, exploitability, and system impact to rank vulnerabilities \cite{shah2022vulnerability, forain2022revs, farris2018vulcon, anjum2020evaluation}. These approaches are generally integrated into broader methodologies or combined with graph-based techniques for more nuanced results. A smaller subset of papers (9) explore statistical methods in combination with other approaches \cite{angelelli2024robust, holm2012empirical, yoon2023assess, nourin2021measuring}. They often played a role in establishing baseline risk metrics or adjusting prioritization rankings through statistical adjustments, or use data-driven approaches like probabilistic techniques to rank vulnerabilities.

\subsubsection{Validation Methods}

Validation approaches vary based on methodological choices. Controlled experiments are the most common, used in 33 papers \cite{kia2024cyber, kurniawan2023automation, mehri2023automated, figueiredo2022exploitability, wen2015novel}, particularly for graph-based and ML/AI approaches. These studies focus on evaluating the accuracy and performance of the prioritization models in a controlled setting. Case studies, presented in 27 papers, typically focusing on real-world scenarios to validate their proposed methodologies \cite{cheimonidis2024dynamic, sharma2023analysis, samtani2022linking, wang2020vulnerability, huang2013novel}. Case studies are often selected for their relevance to the systems or networks under consideration, providing valuable insight into the applicability of the techniques in practice. Simulations, appearing in 13 studies, offer an environment to test theoretical models, though their reliance on synthetic data may limit generalizability \cite{zeng2023illation, yan2022cyber, keskin2021scoring, wang2020bayesian, srivastava2013modeling}. Interviews, appearing in 6 papers, represent a smaller effort to capture expert opinions and contextual insights \cite{jiang2023model, samtani2022linking, longueira2022novel, almukaynizi2017proactive, holm2012empirical, anuar2011risk}.

Validation approaches vary based on methodological choices. Controlled experiments are the most common, used in 33 papers \cite{kia2024cyber, kurniawan2023automation, mehri2023automated, figueiredo2022exploitability, wen2015novel}, particularly for graph-based and ML/AI approaches. These studies focus on evaluating the accuracy and performance of the prioritization models in a controlled setting. Case studies, presented in 27 papers, typically focusing on real-world scenarios to validate their proposed methodologies \cite{cheimonidis2024dynamic, sharma2023analysis, samtani2022linking, wang2020vulnerability, huang2013novel}. Case studies are often selected for their relevance to the systems or networks under consideration, providing valuable insight into the applicability of the techniques in practice. Simulations, appearing in 13 studies, offer an environment to test theoretical models, though their reliance on synthetic data may limit generalizability \cite{zeng2023illation, yan2022cyber, keskin2021scoring, wang2020bayesian, srivastava2013modeling}. Interviews, appearing in 6 papers, represent a smaller effort to capture expert opinions and contextual insights \cite{jiang2023model, samtani2022linking, longueira2022novel, almukaynizi2017proactive, holm2012empirical, anuar2011risk}.

Table \ref{tab:Methodology} in the Appendix summarizes the methodologies and validation approaches used across studies. Validation trends for specific methodologies are as follows.

\begin{itemize}
    \item Graph-based methods partially utilize controlled experiments (12 studies), often using synthetic data or isolated models to test vulnerability propagation. Case studies (11 papers) provide real-world validation, such as network architectures or interconnected devices. Simulations are the least common (7 papers), likely due to their limitations in replicating real-world complexity.
    \item ML-based methods heavily utilize experiments (22 out of 24 studies) which test model accuracy through training and testing on known data splits, such as historical vulnerability datasets. Case studies (6 studies) are also utilized to validate usefulness of the proposed approaches.
    \item Validation of multi-objective optimization methods shows a mixed pattern across experiments (6 out of 10 papers), case studies (5 papers), and simulations (4 papers). Experiments test the methods' ability to balance trade-offs in controlled settings, while case studies provide real-world validation of operational constraints.
    \item Rule-based systems rely on controlled experiments (14 studies), reflecting their dependence on isolated testing. Case studies ( studies) are used to test predefined rules and domain knowledge.
    \item Statistical methods use controlled experiments (3 studies) and case studies (3 studies) to test quantitative models using empirical data.
\end{itemize}

Despite the prevalence of controlled experiments and real-world case studies, crucial details—such as host configurations and network architectures—are often omitted, limiting reproducibility and applicability.

\subsection{Data Quality Challenges}

Table \ref{tab:DataSourceSeverityMetrics} (Appendix) summarizes the data sources used in the reviewed papers. Most studies rely on standard vulnerability databases such as CVE/NVD (70 out of 82) \cite{kia2024cyber, wang2022automotive, nourin2021measuring} and Shodan \cite{angelelli2024robust, reyes2022environment} as foundational inputs. For exploitability-oriented assessments, additional sources include exploit databases (e.g., ExploitDB \cite{samtani2022linking, yoon2023vulnerability}), security reports (e.g., FireEye APT reports \cite{zeng2021licality}, Symantec attack signatures \cite{almukaynizi2017proactive, chen2019using}), and collaborative platforms (e.g., ICS-CERT \cite{yoon2023vulnerability}). Emerging sources, such as social media (Twitter \cite{seker2023xvrs, jacobs2023enhancing}, GitHub \cite{yoon2023vulnerability}), offer real-time insights into exploit announcements. Domain-specific studies, particularly for ICS and automotive systems, incorporate system-specific data to provide contextualized risk assessments \cite{hore2023deep, li2023security, cheimonidis2024dynamic}. 

Vulnerability databases, such as CVE and NVD, are foundational for vulnerability prioritization frameworks. However, their widespread reliance exposes limitations, including data inconsistencies, incompleteness, and delays, which hinder effective and comprehensive vulnerability prioritization. Over-reliance on a single source also limits coverage and reduces adaptability to rapidly evolving threats. For instance, \cite{dong2019towards} finds discrepancies in software version vulnerabilities between CVE and NVD, with only a fraction of entries matching accurately. Similarly, \cite{woo2021v0finder} reveals issues with incorrect or inconsistent software names and versions, emphasizing the need for identifying original vulnerable software. \cite{li2023anatomy} criticizes existing databases for lacking detailed metadata and contextual information, limiting their capacity to support advanced analytical tools effectively. These gaps are particularly problematic for frameworks requiring asset-specific configurations or real-time exploit maturity assessments. Emerging approaches attempt to address these shortcomings. For example, Hong et al. \cite{hong2022xvdb} introduced an enhanced database construction method that correlates NVD entries with additional sources like GitHub, issue trackers (e.g., Bugzilla), and Q\&A sites (e.g., Stack Overflow) to augment data scope and completeness.

Another dimension of data quality is the potential for biases, such as the over-reporting of critical vulnerabilities due to incentives for attention or funding \cite{croft2022investigation}. These biases skew prioritization efforts, diverting resources from vulnerabilities that may pose greater operational risks in specific contexts.

Addressing these challenges requires advancing data aggregation techniques, improving the timeliness and consistencies of vulnerability feeds, and incorporating mechanisms to evaluate and mitigate data biases. By enhancing data quality, prioritization frameworks can deliver more accurate and actionable insights, better aligning with the needs of both researchers and practitioners.

\subsection{Integration of Standards and Compliance}

In our analysis of 82 papers, several key standards (including CVSS, CPE, CWE, CWSS, EPSS, and SCAP) emerged as essential to vulnerability prioritization methodologies. These standards guide the evaluation of vulnerabilities and form the foundation for risk metrics. 

CVSS is the most widely used standard, serving as a baseline for severity assessment and risk prioritization. Its dominance reflects its simplicity and broad applicability across domains. CWE \cite{yin2023empowering, kurniawan2023automation, nourin2021measuring} provides a hierarchical categorization of software weaknesses, that enables more structured vulnerability assessments and aligns vulnerabilities with root causes. However, standards like CVSS and CWE are inherently static, offering limited adaptability to real-time changes, such as active exploitation trends or system-specific contexts. This reduces their effectiveness in dynamic environments. Developing extensions to static standards like CVSS and CWE that account for real-time exploitability, asset criticality, and operational environments is crucial \cite{elder2024survey}. For example, integrating EPSS predictions with CVSS scores can bridge the gap between severity and exploitation likelihood. CPE supports system and software identification, as observed in studies focusing on system-specific analyses \cite{chatzipoulidis2015information}. Its value lies in mapping vulnerabilities to specific software configurations. SCAP is critical for automating vulnerability management and ensuring policy compliance, as highlighted in \cite{chatzipoulidis2015information}. However, its adoption remains limited to studies explicitly addressing automated workflows.

Many frameworks overlook aligning standard-based prioritization with compliance regimes (e.g., PCI DSS, HIPAA, NERC CIP). These regulations impose SLOs/SLA deadlines for remediation, which often conflict with risk-based approaches. Future research should explore adaptive frameworks that incorporate standards like CVSS, CPE, and SCAP alongside compliance-driven metrics (e.g., SLA deadlines). This integration can align risk-based and regulatory priorities for more effective vulnerability management.

\subsection{Adaptive, Explainable, and Scalable Vulnerability Prioritization}

To overcome the limitations of static models, there is a clear shift toward the development of adaptive, context-aware metrics. These metrics incorporate environmental factors such as system configurations, network topology, asset criticality, and operational impact, resulting in more accurate and timely risk assessments that better reflect real-world conditions. 

Many advanced vulnerability prioritization models, particularly ML-based methods, achieve high performance but often lack explainability, functioning as black-box systems \cite{nadeem2023sok}. This limited interpretability reduces trust among cybersecurity practitioners who require clear justifications for risk scores to make actionable decisions. Furthermore, there is often a trade-off between fine-grained analysis and computational efficiency, as highly granular models demand greater resources, impacting scalability. Formal methods and symbolic analysis present promising avenues for addressing these challenges. For example, MulVAL \cite{tayouri2023survey} leverages logic-based attack graph generation to systematically infer exploit chains. Similarly, techniques like Automatic Exploit Generation (AEG) \cite{avgerinos2014automatic} can provide concrete examples of exploit presence, enhancing explainability by demonstrating real-world exploitability. Additionally, quantitative robustness methods \cite{girol2024quantitative, girol2021not} can improve the replicability and reliability of exploit assessments, enabling finer-grained analyses of vulnerabilities under varied conditions. These techniques not only offer deeper insights into exploit behavior but also bridge gaps between interpretability and model performance.

As systems grow in complexity and scale, vulnerability prioritization frameworks must process large datasets and intricate dependencies efficiently. Scalability challenges are compounded by the need for automation, particularly in large-scale networks or real-time prioritization settings \cite{sabur2022toward}. Scalability can be improved through hierarchical frameworks that filter vulnerabilities at a high level while focusing fine-grained analyses on critical assets. Automated tools and standardized vulnerability feeds can streamline workflows while maintaining performance.

\subsection{Industrial Trends}

We analyze several widely adopted industrial solutions (i.e., Tenable One platform \cite{Tenable}, Qualys VMDR \cite{Qualys}, Skybox Security VTM \cite{Skybox}, and Claroty xDome \cite{Claroty}) to elucidate current industrial trends in vulnerability prioritization. 

\textbf{Scalability for Large, Dynamic Environments:} The proliferation of cloud infrastructures, IoT devices, and hybrid IT-OT environments has necessitated scalable vulnerability prioritization solutions. Industry-leading platforms such as Tenable One \cite{Tenable} and Qualys VMDR \cite{Qualys} process high volumes of data across expansive infrastructures while maintaining optimal performance. A notable industry shift is the integration of managed security services (MSSPs), as organizations increasingly rely on third-party providers for vulnerability management, rather than deploying in-house solutions. The MSSP model enables continuous monitoring with minimal internal effort, a trend expected to grow as security providers transition to SaaS-based commercial solutions \cite{TenableMSSP}. Additionally, Claroty xDome \cite{Claroty} and Skybox Security VTM \cite{Skybox} demonstrate adaptability in securing both IT and OT environments, emphasizing automated exposure management rather than traditional manual assessments.

\textbf{Increased Granularity in Risk Assessments:} Tools such as Skybox Security VTM \cite{Skybox} offer detailed vulnerability analyses, encompassing individual attack vectors, asset configurations, and system inter-dependencies. This shift towards higher granularity is particularly significant in sectors like OT and manufacturing, where precise risk identification is critical for maintaining operational safety and security. Fine-grained analysis also helps prioritize vulnerabilities that pose the greatest risk to specific components within a larger system, thereby providing more actionable intelligence for security teams. Additionally, Skybox Security offers virtual patching options, enabling vulnerability prioritization without operational disruption.

\textbf{Adoption of Context-Aware and Multi-Domain Metrics:} A paradigm shift is observable in the development of context-aware and multi-domain metrics for vulnerability prioritization. These metrics consider factors such as asset criticality, operational impact and network topology. For instance, Claroty xDome \cite{Claroty} provides comprehensive risk assessments that incorporate device characteristics, compensating controls, and their position within broader industrial networks. This trend reflects the growing recognition that a vulnerability’s impact can vary significantly depending on its environment, and prioritization models must account for this variability.

\textbf{Integration of AI and ML:} The adoption of AI-driven threat exposure management is gaining traction in industry, surpassing traditional signature-based vulnerability detection. Qualys TotalAI \cite{QualysTotalAI} utilizes LLMs to assess vulnerabilities in AI workloads, detecting data leaks, injection vulnerabilities, and model theft—a capability not widely explored in academic studies. Similarly, Tenable AI Aware \cite{TenableAI} employs network monitoring agents to proactively detect AI-related security risks. Such industrial implementations focus on real-time threat exposure mapping, integrating AI into managed detection \& response workflows to continuously refine attack surface visibility.

%\begin{tcolorbox}[colback=gray!10!white, colframe=blue!75!black, left=2mm, right=0mm, top=1mm, bottom=1mm, boxrule=0pt, sharp corners, before skip=10pt, after skip=10pt, coltitle=black] \textbf{Observation:} The growing emphasis on explainability, granularity, scalability and generalizability, along with the integration of ML techniques and real-time monitoring, reflects the growing demand for advanced and efficient vulnerability prioritization solutions across diverse sectors. \end{tcolorbox}

\section{Conclusion} \label{sec:Conclusion}

This paper presents a comprehensive systematization of cybersecurity risk metrics, offering a novel taxonomy of vulnerability prioritization metrics and identifying key gaps in the existing literature. 

A quantitative analysis of vulnerability prioritization metrics highlights a dominance of impact metrics (60/82 studies) but reveals gaps in operational and predictive assessments. Contextual metrics (48 studies) and exploitability metrics (40 studies) are gaining traction, yet their integration remains limited, with only 22 studies combining them. Methodologically, graph-based (27 studies) and ML/AI-driven approaches (24 studies) are balanced against expert-driven rule-based systems (24 studies), while multi-objective optimization and statistical methods remain underexplored. Validation trends indicate a preference for controlled experiments (33 studies) and case studies (27 studies), though critical details on system configurations are often lacking, limiting reproducibility. These findings emphasize the need for more holistic, multidimensional frameworks integrating operational, predictive, and system-wide risk considerations.

Our analysis highlights the growing need for adaptive, scalable, and context-aware metrics that integrate real-time threat intelligence and dynamically adjust to evolving threats. Existing approaches often rely on static models that struggle to keep pace with the rapidly changing cyber landscape. Future research should focus on developing scalable, automated solutions capable of handling the increasing complexity of modern systems, particularly through adversarial intelligence and dynamic prioritization techniques. Additionally, a critical gap remains in the explainability of AI-driven models, as lack of transparency continues to hinder their adoption in operational settings. While many studies focus on individual systems or isolated vulnerabilities, holistic approaches that account for inter-dependencies across systems and networks are necessary for more effective risk management. Addressing these challenges will be essential for advancing the next generation of vulnerability prioritization frameworks that are not only technically robust but also practically applicable across various domains and industries.

\bibliographystyle{ACM-Reference-Format}
\bibliography{VulRankSurvey}

%%
%% If your work has an appendix, this is the place to put it.
\appendix
\section{Appendix}

Table \ref{tab:DataSourceSeverityMetrics} summarizes the data sources and severity metrics utilized in the reviewed papers. 

Table \ref{tab:methodologies_metrics} correlates the utilized methodologies with the applied risk metrics in the reviewed papers.

Table \ref{tab:Methodology} summarizes the methodologies and validation methods adopted in the reviewed papers.

\begin{table}[h]
    \scriptsize
\renewcommand{\arraystretch}{0.8}
    \centering
    \caption{Data Source and Severity Metric Coverage by Reviewed Paper}
    \label{tab:DataSourceSeverityMetrics}
    \rowcolors{4}{gray!15}{white}
    \begin{tabular}{|l|c|c|c|c|c|c|c|c|c|}
\toprule
\multirow{2}{*}{\textbf{Paper}} & \multicolumn{4}{|c|}{\textbf{Data Sources}} & \multicolumn{5}{c|}{\textbf{Severity Metrics}} \\
\cmidrule(lr){2-5} \cmidrule(lr){6-10}
& \textbf{Standard} & \textbf{Exploit} & \textbf{Social} & \textbf{System} 
& \textbf{Impact} & \textbf{Exploitability} & \textbf{Contextual} 
& \textbf{Predictive} & \textbf{Aggregation} \\
\midrule

        \cite{kia2024cyber} & \textbullet & & \textbullet & & \textbullet & \textbullet & & \textbullet &  \\
        \cite{angelelli2024robust} & \textbullet & \textbullet & & & \textbullet & \textbullet & \textbullet & \textbullet & \\
        \cite{zeng2023illation} & \textbullet & \textbullet & & \textbullet & \textbullet & \textbullet & \textbullet & \textbullet & \textbullet \\
        \cite{cheimonidis2024dynamic} & \textbullet & & & \textbullet & \textbullet & \textbullet & \textbullet & & \textbullet  \\
        \cite{iannone2024early} & \textbullet & \textbullet & & & & \textbullet & & \textbullet & \\
        \cite{seker2023xvrs} & \textbullet & & \textbullet & & \textbullet & \textbullet & & & \\
        \cite{hore2023deep} & \textbullet & \textbullet & & \textbullet & \textbullet & \textbullet & \textbullet & \textbullet & \\
        \cite{yin2023empowering} & \textbullet & \textbullet & & & & \textbullet & & & \\
        \cite{li2023security} & & & & \textbullet & \textbullet & & &  & \textbullet \\
        \cite{kurniawan2023automation} & \textbullet & & & \textbullet & \textbullet & \textbullet & \textbullet & & \\
        \cite{sharma2023analysis} & \textbullet & & & \textbullet & \textbullet & \textbullet & \textbullet & & \textbullet \\
        \cite{yoon2023vulnerability} & \textbullet & \textbullet & \textbullet & & & \textbullet & & & \\
        \cite{wang2023critical} & \textbullet & & & \textbullet & & \textbullet & \textbullet & & \\
        \cite{cheng2023network} & \textbullet & & & \textbullet & \textbullet & & \textbullet & & \\
        \cite{hu2023cost} & \textbullet & & & \textbullet & \textbullet & & \textbullet & & \textbullet \\
        \cite{wang2022automotive} & \textbullet & & & & \textbullet & & \textbullet & & \\
        \cite{mehri2023automated} & \textbullet & & & \textbullet & & & \textbullet &  & \\
        \cite{jacobs2023enhancing} & \textbullet & \textbullet & \textbullet & & & \textbullet & & \textbullet & \\
        \cite{jiang2023model} & \textbullet & & & \textbullet & \textbullet & & \textbullet & & \\
        \cite{oser2022risk} & \textbullet & & & \textbullet & \textbullet & & & \textbullet & \\
        \cite{yan2022cyber} & \textbullet & & & \textbullet & & \textbullet & \textbullet & & \\
        \cite{shah2022vulnerability} & \textbullet & & & & \textbullet & & & & \textbullet \\
        \cite{hore2022towards} & \textbullet & & & & \textbullet & & \textbullet & \textbullet &  \textbullet \\
        \cite{forain2022revs} & \textbullet & \textbullet & & & \textbullet & \textbullet & & & \\
        \cite{samtani2022linking} & \textbullet & \textbullet & \textbullet & & \textbullet & \textbullet & \textbullet & \textbullet & \\
        \cite{almazrouei2022internet} & \textbullet & & & \textbullet & \textbullet & \textbullet & \textbullet & & \textbullet \\
        \cite{costa2022challenges} & \textbullet & \textbullet & & & \textbullet & \textbullet & & & \\
        \cite{jung2022cavp} & \textbullet & & & \textbullet & \textbullet & \textbullet & \textbullet & & \textbullet \\
        \cite{yadav2022smartpatch} & \textbullet & & & \textbullet & \textbullet & & \textbullet & & \\
        \cite{ahmadi2022automated} & \textbullet & & & \textbullet & \textbullet & & \textbullet & & \\
        \cite{figueiredo2022exploitability} & \textbullet & \textbullet & \textbullet & &  & \textbullet &   & \textbullet & \\
        \cite{colombelli2022multi} & & & & \textbullet & \textbullet  & & \textbullet & & \\
        \cite{reyes2022environment} & \textbullet & & & \textbullet & \textbullet & \textbullet & \textbullet & & \\
        \cite{longueira2022novel} & \textbullet & & & & \textbullet & & \textbullet & \textbullet &  \\
        \cite{longueira2022gotta} & \textbullet & \textbullet & & & \textbullet & \textbullet & \textbullet & &  \\
        \cite{zeng2021licality} & \textbullet & \textbullet & \textbullet & &   & \textbullet & \textbullet & \textbullet & \\
        \cite{walkowski2021vulnerability} & \textbullet & & & \textbullet & \textbullet & & \textbullet &   & \\
        \cite{nourin2021measuring} & \textbullet & & & & \textbullet &  & &  & \\
        \cite{northern2021vercasm} & \textbullet & & & & \textbullet & & \textbullet & & \\
        \cite{mahmood2021prioritizing} & & & \textbullet & &  & & \textbullet  & & \\
        \cite{keskin2021scoring} & \textbullet & & & \textbullet & \textbullet & & \textbullet & & \\
        \cite{pecl2021manager} & \textbullet & & & & \textbullet & \textbullet & \textbullet & & \\
        \cite{walkowski2021automatic} & \textbullet & & & \textbullet & \textbullet & & \textbullet &  & \\
        \cite{walkowski2020efficient} & \textbullet & & & \textbullet & \textbullet & & \textbullet & & \\
        \cite{fang2020fastembed} & \textbullet & \textbullet & & & \textbullet & \textbullet & & \textbullet & \\
        \cite{roumani2020examining} & \textbullet & \textbullet & & & &  \textbullet & & & \\
        \cite{wang2020bayesian} & \textbullet & & & &  & & & & \textbullet  \\
        \cite{anjum2020evaluation} & \textbullet & & & &   & \textbullet & &  & \\
        \cite{ani2020vulnerability} & \textbullet & & & &  \textbullet & & \textbullet & & \\
        \cite{yadav2020iot} & \textbullet & & & \textbullet & \textbullet   & & \textbullet & & \\
        \cite{anjum2020framework} & \textbullet & & & &  & & &  & \textbullet \\
        \cite{gourisetti2019cybersecurity} & & & & & & & & & \textbullet \\
        \cite{zhang2020dynamic} & \textbullet & & & \textbullet &  \textbullet & \textbullet & & \textbullet & \\
        \cite{wang2020vulnerability} & \textbullet & & & & \textbullet & \textbullet& \textbullet & & \\
        \cite{dissanayaka2020vulnerability} & \textbullet & \textbullet & & & \textbullet  &   & &  & \\
        \cite{liu2019study} & & & & \textbullet & \textbullet & & \textbullet &  & \textbullet \\
        \cite{chen2019vase} & \textbullet & & \textbullet & & \textbullet & \textbullet & & \textbullet & \\
        \cite{alperin2019risk} & \textbullet & \textbullet & & & \textbullet & \textbullet & & \textbullet & \\
        \cite{zou2019autocvss} & \textbullet & \textbullet & & \textbullet &  \textbullet & & & & \\
        \cite{yadav2019patchrank} & \textbullet & & \textbullet & \textbullet & \textbullet & \textbullet & \textbullet & & \\
        \cite{angelini2018vulnus} & \textbullet & & & \textbullet & \textbullet & \textbullet & \textbullet & & \\
        \cite{chen2019using} & \textbullet & \textbullet & \textbullet & &   & \textbullet & & & \\
        \cite{le2019automated} & \textbullet & & & & \textbullet & & & & \\
        \cite{farris2018vulcon} & \textbullet & & & \textbullet & \textbullet & & \textbullet &   & \textbullet \\
        \cite{tatarinova2018avia} & \textbullet & & & & & \textbullet & \textbullet &  & \textbullet \\
        \cite{khaledian2018power} & & & & \textbullet &  &  & \textbullet & & \\
        \cite{falco2018iiot} & \textbullet & \textbullet & \textbullet & & \textbullet  & \textbullet & &  & \\
        \cite{garg2018empirical} & \textbullet & & & \textbullet & \textbullet & & \textbullet & & \\
        \cite{almukaynizi2017proactive} & \textbullet & \textbullet & \textbullet & & \textbullet & & & & \\
        \cite{chatzipoulidis2015information} & \textbullet & & & \textbullet & \textbullet & & & \textbullet & \\
        \cite{wen2015novel} & \textbullet & \textbullet & & & \textbullet & & & & \\
        \cite{ghosh2015netsecuritas} & & & & \textbullet & & \textbullet & \textbullet & & \textbullet \\
        \cite{bulbul2014cyber} & & & & \textbullet & & \textbullet & \textbullet & & \textbullet \\
        \cite{ouedraogo2013towards} & \textbullet & & & \textbullet & \textbullet & & \textbullet & & \textbullet \\
        \cite{srivastava2013modeling} & & & & \textbullet &  & & \textbullet & & \\
        \cite{huang2013novel} & \textbullet & & & & \textbullet & & & & \\
        \cite{anuar2013incident} & \textbullet & & & \textbullet & \textbullet & & \textbullet & &   \\
        \cite{holm2012empirical} & \textbullet & & & \textbullet & \textbullet & & &  & \textbullet \\
        \cite{anuar2011risk} & \textbullet & & & \textbullet & \textbullet & & \textbullet & &  \\
        \cite{sun2011automatic} & \textbullet & & & \textbullet & \textbullet & & & & \textbullet \\
        \cite{fruhwirth2009improving} & \textbullet & & & & \textbullet & \textbullet & & & \\
        \cite{frei2006large} & \textbullet & & & & & \textbullet & \textbullet & &  \\
        \bottomrule
    \end{tabular}
\end{table}

\begin{table}[h]
    \scriptsize
\renewcommand{\arraystretch}{0.8}
\centering
\caption{Summary of methodologies and metrics for reviewed papers.}
\label{tab:methodologies_metrics}
    \rowcolors{4}{gray!15}{white}
\begin{tabular}{|l|c|c|c|c|c|c|c|c|c|c|}
\hline
\multirow{2}{*}{\textbf{Paper}} & \multicolumn{5}{c|}{\textbf{Methodologies}} & \multicolumn{5}{c|}{\textbf{Metrics}} \\ \cline{2-11} 
                         & \textbf{Graph} & \textbf{ML} & \textbf{Multi-Objective} & \textbf{Rule} & \textbf{Statistical} & \textbf{Impact} & \textbf{Exploitability} & \textbf{Contextual} 
& \textbf{Predictive} & \textbf{Aggregation} \\ \hline
\cite{kia2024cyber} &  & \textbullet &  &  &  & \textbullet & \textbullet &  & \textbullet &  \\
\cite{angelelli2024robust} &  &  &  &  & \textbullet & \textbullet & \textbullet & \textbullet & \textbullet &  \\
\cite{zeng2023illation} &  & \textbullet &  & \textbullet &  & \textbullet & \textbullet & \textbullet & \textbullet & \textbullet \\
\cite{cheimonidis2024dynamic} & \textbullet &  &  & \textbullet &  & \textbullet & \textbullet & \textbullet &  & \textbullet \\
\cite{iannone2024early} &  & \textbullet &  &  &  &  & \textbullet &  & \textbullet &  \\
\cite{seker2023xvrs} &  & \textbullet &  &  &  & \textbullet & \textbullet &  &  &  \\
\cite{hore2023deep} &  & \textbullet &  &  &  & \textbullet & \textbullet & \textbullet & \textbullet &  \\
\cite{yin2023empowering} & \textbullet & \textbullet &  &  &  &  & \textbullet &  &  &  \\
\cite{li2023security} & \textbullet &  &  & \textbullet &  & \textbullet &  &  &  & \textbullet \\
\cite{kurniawan2023automation} &  &  &  & \textbullet &  & \textbullet & \textbullet & \textbullet &  &  \\
\cite{sharma2023analysis} &  &  &  & \textbullet &  & \textbullet & \textbullet & \textbullet &  & \textbullet \\
\cite{yoon2023vulnerability} &  &  &  &  & \textbullet &  & \textbullet &  &  &  \\
\cite{wang2023critical} & \textbullet &  &  &  &  &  & \textbullet & \textbullet &  &  \\
\cite{cheng2023network} & \textbullet &  &  &  &  & \textbullet &  & \textbullet &  &  \\
\cite{hu2023cost} &  &  &  & \textbullet &  & \textbullet &  & \textbullet &  & \textbullet \\
\cite{mehri2023automated} &  &  &  & \textbullet &  &  &  & \textbullet &  &  \\
\cite{jacobs2023enhancing} &  & \textbullet &  &  &  &  & \textbullet &  & \textbullet &  \\
\cite{jiang2023model} & \textbullet &  &  & \textbullet &  & \textbullet &  & \textbullet &  &  \\
\cite{oser2022risk} &  &  &  & \textbullet &  & \textbullet &  &  & \textbullet &  \\
\cite{yan2022cyber} & \textbullet &  &  &  &  &  & \textbullet & \textbullet &  &  \\
\cite{shah2022vulnerability} &  &  & \textbullet &  &  & \textbullet &  &  &  & \textbullet \\
\cite{hore2022towards} &  & \textbullet &  &  &  & \textbullet &  & \textbullet & \textbullet & \textbullet \\
\cite{forain2022revs} &  &   & \textbullet &   &   & \textbullet & \textbullet &  &   &   \\
\cite{samtani2022linking} &  & \textbullet &  &  &  & \textbullet & \textbullet & \textbullet & \textbullet &  \\
\cite{almazrouei2022internet} & \textbullet &  &  &  &  & \textbullet & \textbullet & \textbullet &  & \textbullet \\
\cite{costa2022challenges} &  & \textbullet &  &  &  & \textbullet & \textbullet &  &  &  \\
\cite{jung2022cavp} &  &  &  & \textbullet &  & \textbullet & \textbullet & \textbullet &  & \textbullet \\
\cite{colombelli2022multi} &  &  & \textbullet &  &  & \textbullet &  & \textbullet &  &  \\
\cite{reyes2022environment} &  &  &  & \textbullet & \textbullet & \textbullet & \textbullet & \textbullet &  &  \\
\cite{longueira2022novel} & \textbullet &  &  &  &  & \textbullet &  & \textbullet & \textbullet &  \\
\cite{zeng2021licality} &  & \textbullet &  &  &  &  & \textbullet & \textbullet & \textbullet &  \\
\cite{walkowski2021vulnerability} &  &  &  & \textbullet &  & \textbullet &  & \textbullet &  &  \\
\cite{nourin2021measuring} &  & \textbullet &  &  & \textbullet & \textbullet &  &  &  &  \\
\cite{northern2021vercasm} &  &  &  & \textbullet &  & \textbullet &  & \textbullet &  &  \\
\cite{mahmood2021prioritizing} & \textbullet &  &  &  &  &  &  & \textbullet &  &  \\
\cite{keskin2021scoring} &  &  &  & \textbullet &  & \textbullet &  & \textbullet &  &  \\
\cite{pecl2021manager} &  &  &  & \textbullet &  & \textbullet & \textbullet & \textbullet &  &  \\
\cite{walkowski2021automatic} &  &  &  & \textbullet &  & \textbullet &  & \textbullet &  &  \\
\cite{walkowski2020efficient} &  & \textbullet &  &  &  & \textbullet &  & \textbullet &  &  \\
\cite{fang2020fastembed} &  & \textbullet &  &  &  & \textbullet & \textbullet &  & \textbullet &  \\
\cite{roumani2020examining} &  &  &  &  & \textbullet &  & \textbullet &  &  &  \\
\cite{wang2020bayesian} & \textbullet &  &  &  &  &  &  &  &  & \textbullet \\
\cite{anjum2020evaluation} &  &  & \textbullet &  &  &  & \textbullet &  &  &  \\
\cite{ani2020vulnerability} & \textbullet &  &  &  &  & \textbullet &  & \textbullet &  &  \\
\cite{yadav2020iot} & \textbullet &  &  &  &  & \textbullet &  & \textbullet &  &  \\
\cite{anjum2020framework} &  &  &  & \textbullet &  &  &  &  &  & \textbullet \\
\cite{gourisetti2019cybersecurity} &  &  & \textbullet &  &  &  &  &  &  & \textbullet \\
\cite{zhang2020dynamic} &  & \textbullet &  &  &  & \textbullet & \textbullet &  & \textbullet &  \\
\cite{wang2020vulnerability} & \textbullet &  &  &  &  & \textbullet & \textbullet & \textbullet &  &  \\
\cite{dissanayaka2020vulnerability} &  &  &  & \textbullet &  & \textbullet &  &  &  &  \\
\cite{liu2019study} &  &  &  & \textbullet &  & \textbullet &  & \textbullet &  & \textbullet \\
\cite{chen2019vase} & \textbullet & \textbullet &  &  &  & \textbullet & \textbullet &  & \textbullet &  \\
\cite{alperin2019risk} &  & \textbullet &  &  &  & \textbullet & \textbullet &  & \textbullet &  \\
\cite{zou2019autocvss} &  &  &  & \textbullet &  & \textbullet &  &  &  &  \\
\cite{yadav2019patchrank} & \textbullet &  &  &  &  & \textbullet & \textbullet & \textbullet &  &  \\
\cite{angelini2018vulnus} & \textbullet &  &  &  &  & \textbullet & \textbullet & \textbullet &  &  \\
\cite{chen2019using} & \textbullet & \textbullet &  &  &  &  & \textbullet &  &  &  \\
\cite{le2019automated} &  & \textbullet &  &  &  & \textbullet &  &  &  &  \\
\cite{farris2018vulcon} &  &  & \textbullet &  &  & \textbullet &  & \textbullet &  & \textbullet \\
\cite{tatarinova2018avia} & \textbullet & \textbullet &  &  &  &  & \textbullet & \textbullet &  & \textbullet \\
\cite{khaledian2018power} & \textbullet &  &  &  &  &  &  &  & \textbullet &  \\
\cite{falco2018iiot} &  &  &  &  & \textbullet & \textbullet & \textbullet &  &  &  \\
\cite{garg2018empirical} & \textbullet &  &  &  &  & \textbullet &  & \textbullet &  &  \\
\cite{almukaynizi2017proactive} &  & \textbullet &  &  &  & \textbullet &  &  &  &  \\
\cite{chatzipoulidis2015information} & \textbullet &  &  &  & \textbullet & \textbullet &  &  & \textbullet &  \\
\cite{wen2015novel} &  & \textbullet &  &  &  & \textbullet &  &  &  &  \\
\cite{bulbul2014cyber} &  &  &  & \textbullet &  &  & \textbullet & \textbullet &  & \textbullet \\
\cite{ouedraogo2013towards} & \textbullet &  &  &  &  &  & \textbullet & \textbullet &  & \textbullet \\
\cite{srivastava2013modeling} & \textbullet &  &  &  &  & \textbullet &  & \textbullet &  & \textbullet \\
\cite{ghosh2015netsecuritas} & \textbullet &  &  &  &  &  &  &  & \textbullet &  \\
\cite{huang2013novel} &  &  & \textbullet &  &  & \textbullet &  &  &  &  \\
\cite{anuar2013incident} &  &  & \textbullet &  &  & \textbullet &  & \textbullet &  &  \\
\cite{holm2012empirical} &  &  &  &  & \textbullet & \textbullet &  &  &  & \textbullet \\
\cite{anuar2011risk} &  &  & \textbullet &  &  & \textbullet &  & \textbullet &  &  \\
\cite{sun2011automatic} &  &  &  & \textbullet &  & \textbullet &  &  &  & \textbullet \\
\cite{fruhwirth2009improving} &  &  &  & \textbullet &  & \textbullet & \textbullet &  &  &  \\
\cite{frei2006large} &  &  &  &  & \textbullet &  & \textbullet & \textbullet &  &  \\
\hline
\end{tabular}
\end{table}

\begin{table}[h]
    \scriptsize
\renewcommand{\arraystretch}{0.8}
    \centering
    \caption{Methodologies and Validation Methods of Reviewed Paper}
    \label{tab:Methodology}
    \rowcolors{4}{gray!15}{white}
\begin{tabular}{|l|c|c|c|c|c|c|c|c|c|}
\hline
        \multirow{2}{*}{\textbf{Paper}} & \multicolumn{5}{|c|}{\textbf{Methodology}} & \multicolumn{4}{c|}{\textbf{Validation Method}} \\
        \cmidrule(lr){2-6} \cmidrule(lr){7-10}
         & \textbf{Graph} & \textbf{ML} & \textbf{Multi-Objective} & \textbf{Rule} & \textbf{Statistical} & \textbf{Case Study} & \textbf{Experiment} & \textbf{Simulation} & \textbf{Interview} \\
\hline
                 \cite{kia2024cyber} &             &      \textbullet &                 &             &             & \textbullet & \textbullet &             &             \\
          \cite{angelelli2024robust} &             &                  &                 &             & \textbullet &             &             & \textbullet &             \\
             \cite{zeng2023illation} &             &      \textbullet &                 & \textbullet &             & \textbullet &             & \textbullet &             \\
       \cite{cheimonidis2024dynamic} & \textbullet &                  &                 & \textbullet &             & \textbullet & \textbullet &             &             \\
             \cite{iannone2024early} &             &      \textbullet &                 &             &             &             & \textbullet &             &             \\
                \cite{seker2023xvrs} &             &      \textbullet &                 &             &             &             & \textbullet &             &             \\
                 \cite{hore2023deep} &             &      \textbullet &                 &             &             &             & \textbullet & \textbullet &             \\
            \cite{yin2023empowering} & \textbullet &      \textbullet &                 &             &             &             & \textbullet &             &             \\
               \cite{li2023security} & \textbullet &                  &                 & \textbullet &             & \textbullet &             &             &             \\
      \cite{kurniawan2023automation} &             &                  &                 & \textbullet &             &             & \textbullet &             &             \\
           \cite{sharma2023analysis} &             &                  &                 & \textbullet &             & \textbullet & \textbullet &             &             \\
        \cite{yoon2023vulnerability} &             &                  &                 &             & \textbullet & \textbullet &             &             &             \\
             \cite{wang2023critical} & \textbullet &                  &                 &             &             &             & \textbullet &             &             \\
             \cite{cheng2023network} & \textbullet &                  &                 &             &             &             & \textbullet &             &             \\
                   \cite{hu2023cost} &             &                  &                 & \textbullet &             &             & \textbullet &             &             \\
           \cite{wang2022automotive} & \textbullet &      \textbullet &                 &             &             &             & \textbullet &             &             \\
           \cite{mehri2023automated} &             &                  &                 & \textbullet &             &             & \textbullet &             &             \\
          \cite{jacobs2023enhancing} &             &      \textbullet &                 &             &             &             & \textbullet &             &             \\
          \cite{jiang2023model} & \textbullet & & & \textbullet & & \textbullet & & & \textbullet \\
                 \cite{oser2022risk} &             &                  &                 & \textbullet &             &             & \textbullet &             &             \\
                 \cite{yan2022cyber} & \textbullet &                  &                 &             &             &             &             & \textbullet &             \\
        \cite{shah2022vulnerability} &             &                  &     \textbullet &             &             &             & \textbullet &             &             \\
              \cite{hore2022towards} &             &      \textbullet &                 &             &             &             & \textbullet &             &             \\
               \cite{forain2022revs} &             &                  &     \textbullet &             &             &             & \textbullet & \textbullet &             \\
           \cite{samtani2022linking} &             &      \textbullet &                 &             &             & \textbullet & \textbullet &             & \textbullet \\
       \cite{almazrouei2022internet} & \textbullet &                  &                 &             &             &             &             &             &             \\
       \cite{costa2022challenges} &             &      \textbullet &                 &             &             &             & \textbullet &             &             \\
                 \cite{jung2022cavp} &             &                  &                 & \textbullet &             & \textbullet &             &             &             \\
          \cite{yadav2022smartpatch} &             &                  &     \textbullet &             &             & \textbullet &             &             &             \\
          \cite{ahmadi2022automated} &             &                  &                 & \textbullet &             &             & \textbullet &             &             \\
 \cite{figueiredo2022exploitability} &             &      \textbullet &                 &             &             &             & \textbullet &             &             \\
          \cite{colombelli2022multi} &             &                  &     \textbullet &             &             &             & \textbullet &             &             \\
         \cite{reyes2022environment} &             &                  &                 & \textbullet & \textbullet &             & \textbullet &             &             \\
           \cite{longueira2022novel} & \textbullet &                  &                 &             &             & \textbullet &             &             & \textbullet \\
           \cite{longueira2022gotta} & \textbullet &                  &                 &             &             & \textbullet &             &             &             \\
             \cite{zeng2021licality} &             &      \textbullet &                 &             &             & \textbullet &             &             &             \\
   \cite{walkowski2021vulnerability} &             &                  &                 & \textbullet &             &             & \textbullet &             &             \\
          \cite{nourin2021measuring} &             &      \textbullet &                 &             & \textbullet &             & \textbullet &             &             \\
          \cite{northern2021vercasm} &             &                  &                 & \textbullet &             &             & \textbullet &             &             \\
      \cite{mahmood2021prioritizing} & \textbullet &                  &                 &             &             & \textbullet &             &             &             \\
            \cite{keskin2021scoring} &             &                  &                 & \textbullet &             &             &             & \textbullet &             \\
              \cite{pecl2021manager} &             &                  &                 & \textbullet &             & \textbullet &             &             &             \\
       \cite{walkowski2021automatic} &             &                  &                 & \textbullet &             &             & \textbullet &             &             \\
       \cite{walkowski2020efficient} &             &      \textbullet &                 &             &             &             & \textbullet &             &             \\
            \cite{fang2020fastembed} &             &      \textbullet &                 &             &             &             & \textbullet &             &             \\
         \cite{roumani2020examining} &             &                  &                 &             & \textbullet &             & \textbullet &             &             \\
             \cite{wang2020bayesian} & \textbullet &                  &                 &             &             &             &             & \textbullet &             \\
          \cite{anjum2020evaluation} &             &                  &     \textbullet &             &             & \textbullet &             &             &             \\
         \cite{ani2020vulnerability} & \textbullet &                  &                 &             &             & \textbullet &             & \textbullet &             \\
                 \cite{yadav2020iot} & \textbullet &                  &                 &             &             &             & \textbullet &             &             \\
           \cite{anjum2020framework} &             &                  &                 & \textbullet &             & \textbullet &             &             &             \\
  \cite{gourisetti2019cybersecurity} &             &                  &     \textbullet &             &             &             & \textbullet &             &             \\
             \cite{zhang2020dynamic} &             &      \textbullet &                 &             &             &             & \textbullet &             &             \\
        \cite{wang2020vulnerability} & \textbullet &                  &                 &             &             & \textbullet &             &             &             \\
 \cite{dissanayaka2020vulnerability} &             &                  &                 & \textbullet &             &             & \textbullet &             &             \\
                 \cite{liu2019study} &             &                  &                 & \textbullet &             & \textbullet &             &             &             \\
                 \cite{chen2019vase} & \textbullet &      \textbullet &                 &             &             &             & \textbullet &             &             \\
              \cite{alperin2019risk} &             &      \textbullet &                 &             &             & \textbullet & \textbullet &             &             \\
              \cite{zou2019autocvss} &             &                  &                 & \textbullet &             &             & \textbullet &             &             \\
           \cite{yadav2019patchrank} & \textbullet &                  &                 &             &             &             & \textbullet & \textbullet &             \\
           \cite{angelini2018vulnus} & \textbullet &                  &                 &             &             &             & \textbullet &             &             \\
                \cite{chen2019using} & \textbullet &      \textbullet &                 &             &             & \textbullet & \textbullet &             &             \\
               \cite{le2019automated} &             &      \textbullet &                 &             &             &             & \textbullet &             &             \\
             \cite{farris2018vulcon} &             &                  &     \textbullet &             &             & \textbullet &             &             &             \\
           \cite{tatarinova2018avia} & \textbullet &      \textbullet &                 &             &             &             & \textbullet &             &             \\
           \cite{khaledian2018power} & \textbullet &                  &                 &             &             & \textbullet &             & \textbullet &             \\
                \cite{falco2018iiot} &             &                  &                 &             & \textbullet &             &             &             &             \\
            \cite{garg2018empirical} & \textbullet &                  &                 &             &             &             & \textbullet &             &             \\
     \cite{almukaynizi2017proactive} &             &      \textbullet &                 &             &             &             & \textbullet &             & \textbullet \\
\cite{chatzipoulidis2015information} & \textbullet &                  &                 &             & \textbullet & \textbullet &             &             &             \\
                 \cite{wen2015novel} &             &      \textbullet &                 &             &             &             & \textbullet &             &             \\
              \cite{bulbul2014cyber} &             &                  &                 & \textbullet &             &             & \textbullet &             &             \\
         \cite{ouedraogo2013towards} & \textbullet &                  &                 &             &             &             &             &             &             \\
       \cite{srivastava2013modeling} & \textbullet &                  &                 &             &             &             &             & \textbullet &             \\
        \cite{ghosh2015netsecuritas} & \textbullet &                  &                 &             &             &             &             & \textbullet &             \\
               \cite{huang2013novel} &             &                  &     \textbullet &             &             & \textbullet &             &             &             \\
            \cite{anuar2013incident} &             &                  &     \textbullet &             &             &             & \textbullet &             &             \\
            \cite{holm2012empirical} &             &                  &                 &             & \textbullet & \textbullet &             &             & \textbullet \\
                \cite{anuar2011risk} &             &                  &     \textbullet &             &             &             & \textbullet &             & \textbullet \\
             \cite{sun2011automatic} &             &                  &                 & \textbullet &             &             &             &             &             \\
       \cite{fruhwirth2009improving} &             &                  &                 & \textbullet &             &             &             & \textbullet &             \\
       \cite{frei2006large}	 &	 &	 &	 &	 & \textbullet	 &	 &	 &	 &	\\																	
\hline
\end{tabular}
\end{table}

%%
%% The next two lines define the bibliography style to be used, and
%% the bibliography file.

\end{document}